\documentclass[12pt,preprint, numberedappendix]{emulateapj}

\usepackage[usenames,graphicx]{}
\usepackage{color}
\usepackage{soul}
\usepackage{bm}

\definecolor{ultramarine}{rgb}{0.07, 0.04, 0.56}

\slugcomment{}

\shorttitle{Time-dependent Pattern Speeds in Barred Galaxies}
\shortauthors{Yu Ting Wu et al.}

\begin{document}


\title{Time-dependent Pattern Speeds in Barred Galaxies}


\author{Yu-Ting Wu\altaffilmark{1,2}, Daniel Pfenniger\altaffilmark{3}  and
  Ronald E. Taam\altaffilmark{2,4}}
\affil{Chile Observatory, National Astronomical Observatory of Japan, Mitaka, Tokyo 181-8588, Japan}
\affil{Institute of Astronomy and Astrophysics, Academia Sinica, Taipei 10617, Taiwan}
\affil{Geneva Observatory, University of Geneva, CH-1290 Sauverny, Switzerland}
\affil{Department of Physics and Astronomy, Northwestern University, 2145 Sheridan Road, Evanston, IL 60208, USA}
\email{yuting.wu@nao.ac.jp}




\begin{abstract}
   Based on a high quality $N$-body simulation of a double bar galaxy model, we investigate the evolution of the bar
    properties, including their size, strength and instantaneous pattern speed derived by using three distinct methods: the Fourier,
    Jacobi integral, and moment of inertia methods.  The interaction of the two bars, which rotate at distinct speeds, primarily
    affects the size, strength and pattern speed of the inner bar.  When the two bars are perpendicular to each other, the size and
    the pattern speed of the inner bar decrease and its strength increases.  The emergence of a strong Fourier $m=1$ mode increases
    the oscillation amplitude of the size, strength and pattern speed of the inner bar.  On the other hand, the characteristics of
    the outer bar are substantially influenced by its adjacent spiral structure.  When the spiral structure disappears, the size of
    the outer bar increases and its strength and pattern speed decrease.  Consequently, the ratio of the pattern speed of the outer
    bar with respect to the inner bar is not constant and increases with time.  Overall, the double bar and disk system displays
    substantial high frequency semi-chaotic fluctuations of the pattern strengths and speeds both in space and time, superposed on the
    slow secular evolution, which invalidates the assumption that the actions of individual stars should be well conserved in barred
    galaxies, such as the Milky Way.
\end{abstract}





\section{Introduction}

Single or double stellar bars are common structures existing in most disk galaxies, playing an important role in the secular
evolution of galaxies driven by internal processes.  More than two thirds of nearby disk galaxies observed in the near-infrared are
considered to have bars, although this fraction is likely higher since bars viewed edge-on are difficult to detect, as illustrated
by the Milky Way, the bar of which was established observationally only in the 1990's. The fraction of galaxies with bars is also
known to depend on the specific properties of galaxies, such as Hubble type (Marinova \& Jogee 2007; Menéndez-Delmestre et al.\
2007). Physically, the presence of bars significantly affects the evolution of a galaxy by redistributing the angular momentum
between its different components; for example, between the disk and halo (Weinberg 1985; Villa-Vargas et al.\ 2009).

In addition, the interaction of two bars can also affect the stellar kinematics of the galaxies, especially for the central region
which is closely related to the inner bar. For example, the $\sigma$-humps (Emsellem et al.\ 2001), which measures the
velocity dispersion has two local maximum along the minor axis of the inner bar, is generated by the existence of vertically thin
bars (Wozniak et al.\ 2003; Du et al.\ 2017a) and can oscillate in strength according to the relative angle between the two
bars (Du et al.\ 2016).

Furthermore, the bars not only affect the stars, but also the gas in the galaxies. Bars are expected to induce gas inflows,
enhancing episodic star formation activity (Aguerri 1999) and/or feeding supermassive black holes in the central regions of galaxies
(Shlosman et al. 1989, Friedli \& Martinet 1993).  However, the existence of the black hole with mass $\sim0.05\%-0.2\%$
of the total stellar mass can destroy the inner bar of a double bar galaxy (Du et al.\ 2017b) as well as in single barred
  galaxies (Hasan \& Norman 1990; Pfenniger \& Norman 1990; Friedli \& Pfenniger 1991; Hasan et al.\ 1993; Norman et al.\
  1996). Therefore, the gas inflows which are driven by the inner bar may be terminated when the black hole grows to $\sim0.1\%$ of
the total stellar mass.

Although bars are expected to affect the evolution of galaxies profoundly, the details of their formation, evolution and
characteristics are still unclear and need to be investigated in more detail. In particular, the evolution of bar sizes, strength
and pattern speeds, as well as the mutual perturbations of double bars and adjacent spirals remain to be better investigated.

Since the 1960's, stellar bars have been assumed as rigidly and steadily rotating structures perturbing a background fixed
gravitational potential (e.g., de Vaucouleurs et al. 1968; Weinberg 1985).  This unchecked assumption is implicit and is used in
most studies on the topic (e.g., Hernquist \& Weinberg 1992; Binney \& Tremaine 2008, Sect. 3.3.2).  However Sellwood \&
  Sparke (1988) first showed that a single bar and its adjacent spiral arms rotate in average at distinct speeds.  When adjacent bar
  or spiral structures rotate with different speeds, their mutual torques cause the intermediate region to be strongly time-dependent
  in any rotating frame. This point was investigated quantitatively by Wu et al. (2016) who demonstrated that bars and spiral arms
   in a double barred galaxy model are actually flexible structures especially in the vicinity of the corotation region. This
results from the fact that the equilibrium points within the corotation region are time-dependent in any rotating frame due to the
mutual interaction with their adjacent structure (bar-bar or bar-spiral structures).  One of the important consequences of the
time-dependent dynamics is that the bars in their entirety do not rotate at a constant pattern speed, especially near their ends
where the pattern speed varies both spatially and temporally.  This implies that we need to be cognizant of the fact that the
instantaneous pattern speed near the corotation region may differ from the instantaneous pattern speed in the other parts of the
bars. From this view, the meaning of the corotation resonance is also challenged since it lies midway between two patterns of
similar strength, but with differing pattern speeds.  At best we could use the term \textit{time-dependent resonance for its
  description}. The significance of the bar Outer Lindblabd resonance (OLR) is even more challenged since it lies in the spiral
region (in the Milky Way close to the Sun) where the bar torque may be weaker than the local spiral torque.

The existing methods to observationally determine pattern speeds in the Milky Way and in external galaxies (e.g., Tremaine \&
  Weinberg 1984) can, to a certain degree, measure instantaneous pattern speeds, whereas the pattern speeds determined in
theoretical studies using N-body simulations (e.g., Sellwood \& Athanassoula 1986) are usually space and time averaged.  To
ensure a more meaningful comparison between theory and observations, it is necessary to develop new methods to measure the local and
instantaneous pattern speeds in numerical simulations, which has been advanced in a companion paper (Pfenniger et al.\ 2018).  In
the present paper, three of these methods (explained in Appendices A--C) are used for analysing a high quality $N$-body
simulation of a double bar disk galaxy model whose potential depends on time in any coordinates.  Following the results in Wu et
al. (2016), it is essential to quantify the different time-dependent characteristics of bars.

The purpose of evolving and analyzing N-body models in this paper is not to describe a fully realistic evolution of a
  galaxy, but to check the degree of time-dependence of different patterns in a collisionless self-gravitating rotating disk similar
  to a galaxy.  Three main properties are studied, the size, strength and instantaneous mode/pattern speeds of the bars.

Our paper is organized as follows.  In \S 2, the galaxy model is described.  The comparison of three methods for determining the
mode/pattern speed and the spatial and temporal evolution of the entire galaxy over 7.8\,Gyr is shown in \S 3. In \S 4 and \S 5, we
focus on the evolution of the inner and outer bar and their interplay.  Finally, we conclude in the last section.

\section{Galaxy Model}

\subsection{The Parameters}

Our initial galaxy model consists of three concentric axisymmetric Miyamoto-Nagai (hereafter MN) components (Miyamoto \& Nagai
  1975), that could be associated with a bulge, a disk, and a halo.  Table~\ref{table:3MN} shows the parameters of these three
components, which are the same as in Wu et al. (2016). To generate double-barred galaxies, our equilibrium initial models should
have (i) a double-peaked rotation curve with equal maxima, one peak corresponding to the disk component and the other to the
  bulge component, (ii) the smaller component is characterized by colder kinematics ($Q \sim 1$), and (iii) the medium component is
described by hotter kinematics ($Q \sim 1.5$), similar to Du et al. (2015). In addition, to maintain a high velocity in the rotation
curve in the spiral region, a spherical component (i.e., a Plummer model) is adopted for the largest MN component. For this
set up, a long-lived double barred galaxy surrounded by a spiral region ensues.

The initial three-dimensional N-body model is constructed using the GalIC code (Yurin \& Springel 2014), modified to generate MN
components. It is evolved using the pure stellar dynamical code gyrfalcON (Dehnen 2000), using the Fast Multipole Methods which has
better momentum conservation than the Barnes-Hut Treecode method. We adopt a total number of equal mass particles
$N = 2 \times 10^7$, which corresponds to a gravitational softening length of $\sim 0.03\,$kpc, in order to reduce the particle
noise and to allow investigation of the inner bar, which has a half-length of $\sim 0.6\,$kpc.  In our model, all components are
live, so that the disk particles are allowed to interact with the halo and bulge particles.  The dynamics is then more realistic than in models which have a fixed halo (e.g., Du et al.\ 2015), thereby suppressing the growth of odd modes.

Our model galaxy is simulated for about 8\,Gyr, which is sufficiently long to study the overall evolution of the two nested bars and
the surrounding spiral arms. The time evolution of the projected density of the double-barred system will be shown in the next
subsection.

\begin{table}[h]   
\begin{center}
  \caption[Parameters $a$, $b$, and $M$ of three Miyamoto-Nagai components.] 
  {Parameters $a$, $b$, and $M$ (as defined in Miyamoto \& Nagai 1975) of three MN components}
  
 \begin{tabular}{lcccc}
 \hline
 \hline
 \multicolumn{2}{l}{Parameter}& Bulge & Disk & Halo \\
 \hline
 Mass $M$ ($10^{10} \, \rm M_{\odot}$) && 1.3496 & 8.6504 &  15.0\\
 Scale length ($a+b$) (kpc)     && 0.50 & 4.50 & 15.0\\
 Scale height $b$ (kpc)     && 0.15 & 0.45 & 15.0\\
 Number of particles   && 1,079,680 & 6,920,320 & 12,000,000\\
 \hline
 \end{tabular}
\label{table:3MN}
\end{center}
\end{table}

\subsection{Time Evolution of the Galaxy}

\begin{figure}[h]
\begin{center}
\includegraphics[trim={2cm 2cm 2cm 2.5cm},clip, scale=0.9]{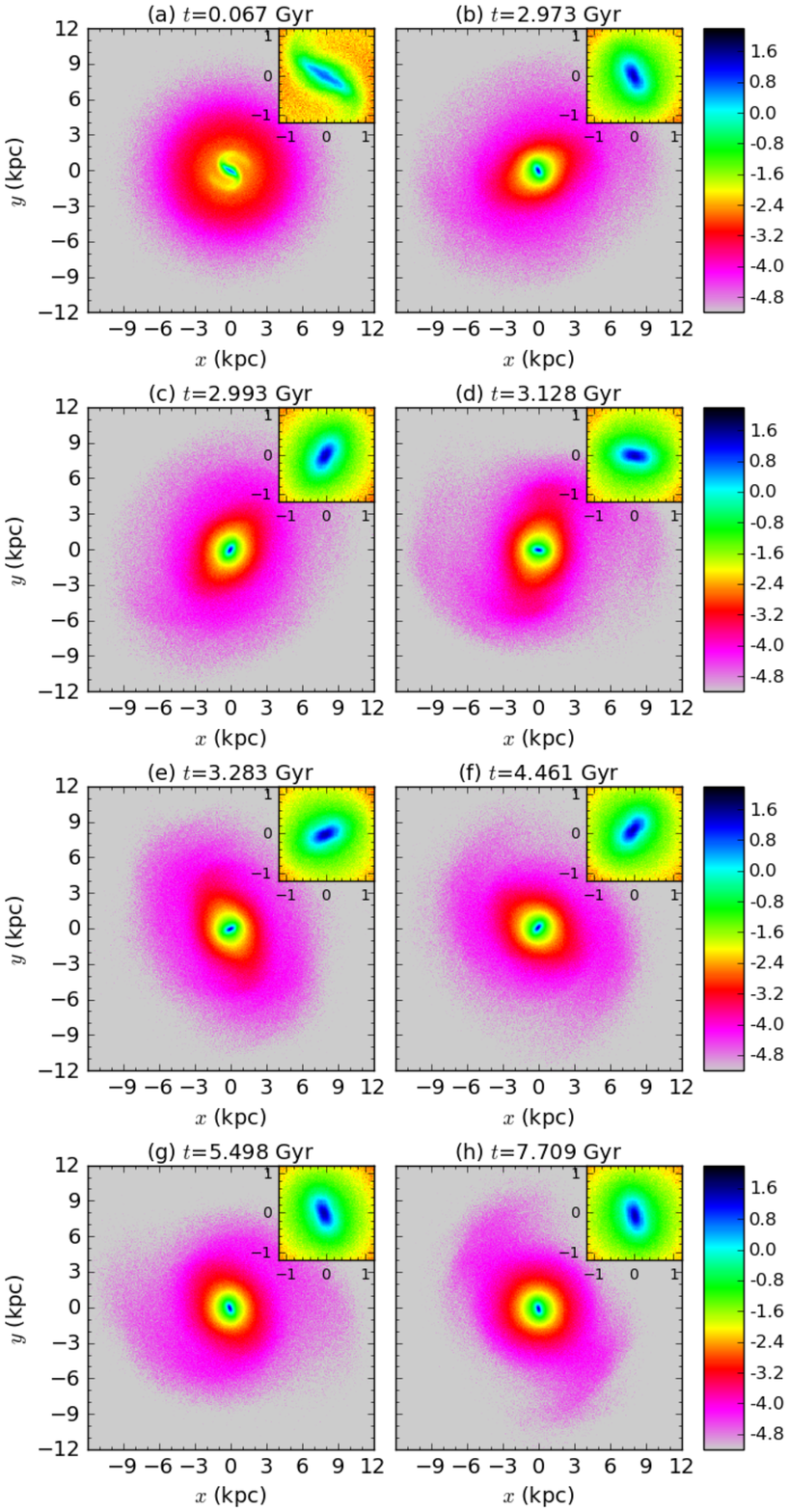}
\caption{Time evolution of the projected density in log-scale at selected times. The inset of each panel shows the projected density
  near the galaxy center.}
\label {fig:model_015_xy_evo}
\end{center}
\end{figure}

Fig.~\ref{fig:model_015_xy_evo} presents the projected density of the double-barred system in log-scale at selected times.  The
inset of each panel shows the projected density near the galaxy center, clearly revealing the inner bar.
  
Starting from the equilibrium initial condition with the bulge, disk and halo components, the inner bar and transient inner spirals
form first due to the colder kinematics near the galaxy center, as shown in panel (a). The inner spirals recur several times until
$t\sim0.5$\,Gyr. Subsequently, the transient outer spirals and the outer bar form. It is worth noticing that in this scenario,
forming the inner bar first and the outer bar later, is not necessarily the only scenario for forming a double barred
galaxy.  A plausible scenario (Shlosman et al. 1989; Friedli \& Martinet 1993) is that the inner bar forms after the outer bar
  when sufficient gas has accumulated as a nuclear disk, and star formation from this disk injects stars with cold kinematics therein,
  triggering a bar instability.

Because the pattern speed of the inner bar and outer bar differ, the phase angle between the two bars varies with time.  Panel
  (b) shows an example at a time when the inner bar and the outer bar are perpendicular to each other.  Panel (c) illustrates
the time when the two bars are aligned.  We note that after $t\sim0.7$\,Gyr, the transient outer spirals recur several times,
instead of forming steady spirals, until the end of the simulation.  For example, as shown in panels (d) and (e), the outer strong
spirals appear at $t=3.128$\,Gyr and then disappear in about 155 Myr.

After $t\sim3.8$\,Gyr, the center of the inner bar starts to clearly shift around the center of the galaxy representing motion
associated with a Fourier $m=1$ mode, an example of which can be seen in the inset of panel (f).  After $t\sim4.8$\,Gyr, the
outer bar becomes oval in shape and cannot be identified well, as shown in panel (g).  However, at the late stage of the simulation,
a longer outer bar with the size $\sim 6$\, kpc and transient spirals recur, as shown in panel (h), lasting until the end of the
simulation.

\begin{figure}[h]
\begin{center}
\includegraphics[trim={0cm 0cm 0cm 0cm},clip, scale=0.6]{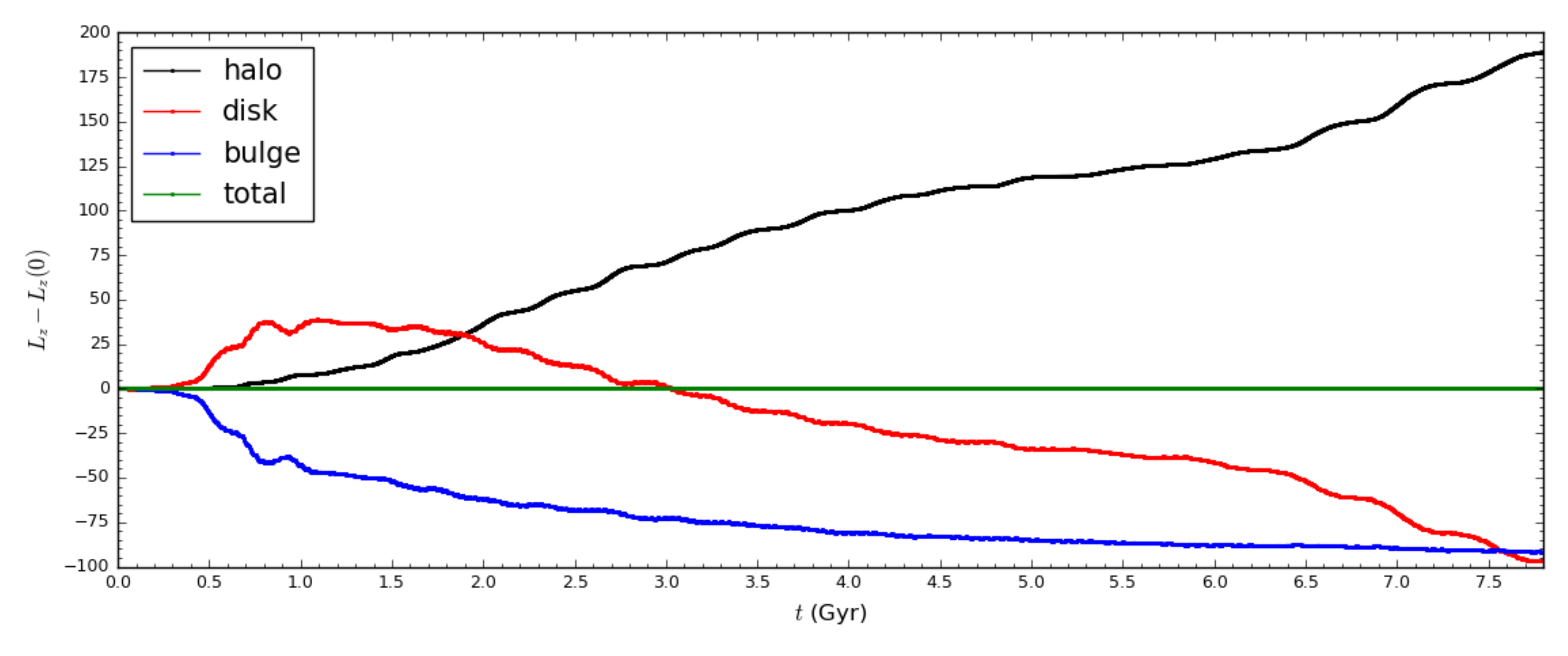}
\caption{Time evolution of the angular momentum change $L_z-L_z(0)$, where $L_z$ is the $z$-component of  the angular momentum and
  $L_z(0)$ is the initial angular momentum. The black, red and blue lines correspond to the halo, disk and bulge components,
  respectively. The green line shows the change of the total $L_z$.}
\label {fig:AM_Lz_evo}
\end{center}
\end{figure}

Fig.~\ref{fig:AM_Lz_evo} shows the angular momentum change $L_z-L_z(0)$ as a function of time, where $L_z$ is the $z$-component of  the angular momentum and $L_z(0)$ is its initial value. The black, red and blue lines correspond to the halo, disk and bulge
components, respectively. The green line shows the total $L_z$, which is well conserved over the full simulation time. 
For 
$t\sim0 - 0.7$\,Gyr, when the inner bar starts to develop, angular momentum is mainly exchanged between the bulge and disk
components, as the red and blue curves mirror each other.  After $t\sim2$\,Gyr, the red and blue lines show that the angular
momentum is exchanged primarily between the disk and halo.  The angular momentum of the halo component increases as the bulge and
disk components lose their angular momentum, as suggested by Debattista \& Sellwood (2000) and Athanassoula (2003). Finally, the
exchange of angular momentum between the disk and the halo is inefficient for $t\sim4.5 - 6.0$\,Gyr as the outer bar becomes weaker.

\section{The Mode/Pattern Speeds and Strengths}

Here, three methods are used to evaluate the rotation speeds of the inner and outer bars: (1) the Fourier method for the {\em mode}
speed of different modes, (2) the Jacobi integral method for the the rotational angular speed of the gravitational potential, and
(3) the moment of inertia method for the bisymmetric mass moment angular speed.  An advantage in using these methods is that
the instantaneous speed is determined rather than the time-averaged speed frequently used in the N-body spiral/bar model
  literature. Hence, the short time-scale variability of the system can be probed.

A brief description of the three methods is given in Appendices A -- C with each method applied to a few snapshots of the
N-body simulation to illustrate their adequacy.  Note that the different methods measure different characteristics of the structure,
giving slightly different results because the structure is actually not time-independent.  No discrepancies occur when the
  structure is rigid and constantly rotating, which has been tested by imposing a solid body rotation to the particles and by
calculating the rotation speed with the described methods.  Using three different methods allows us to better grasp the weak or strong points of each method and to better understand the double-barred system.  In the following, we compare the results using the
different methods in \S 3.1, and show the time evolution of the mode/pattern speed and the bar strength over the full simulation
time in \S 3.2.

\subsection{Comparison of the Different Methods}

\begin{center}
\begin{figure}[h]
  \epsscale{1.1}
 \plotone{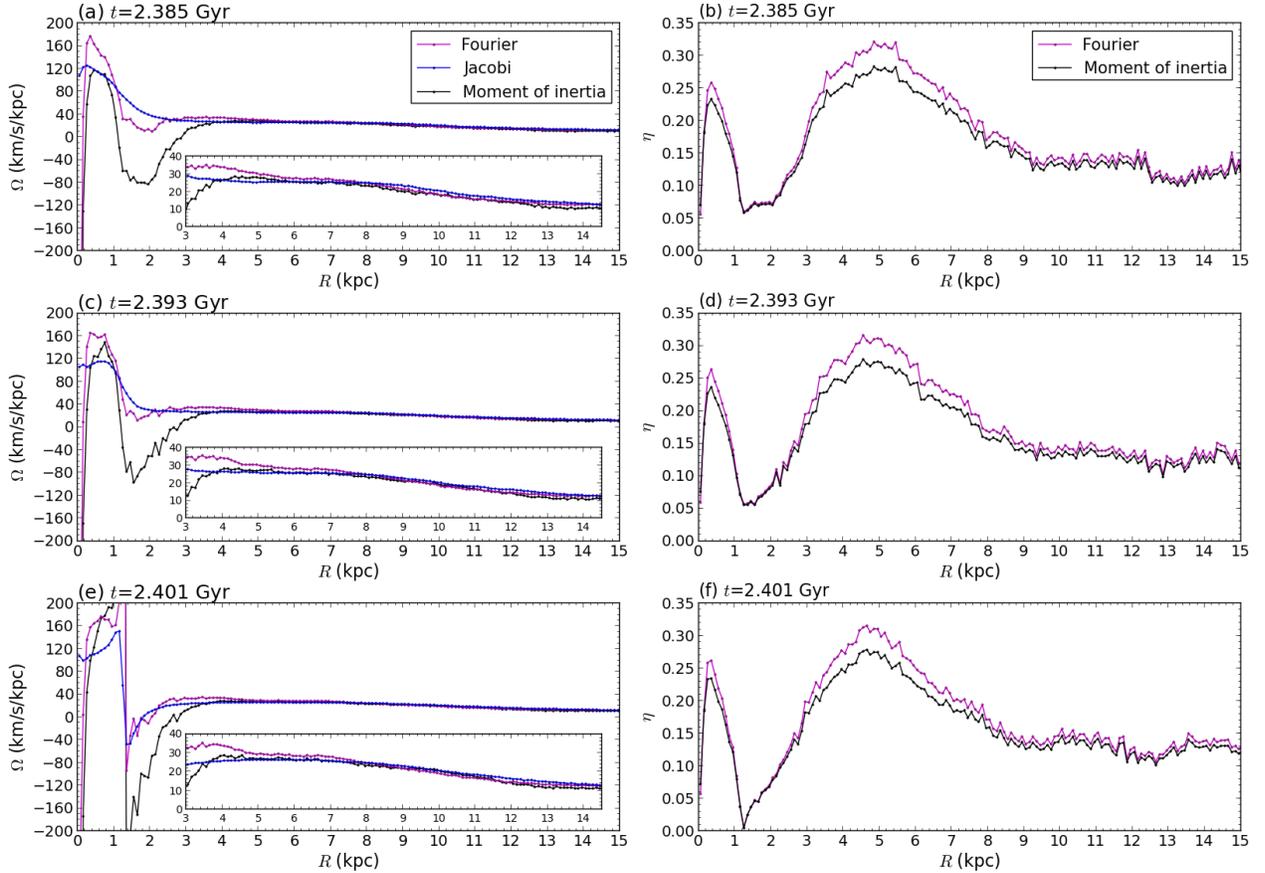}
 \caption{Comparison of the radial  mode/pattern speed profile (left panels) and the strength (right panels) at the selected times
   $t=2.385$, 2.393, and 2.401\,Gyr.  The magenta, blue and black lines denote determinations based on the Fourier method, the
   Jacobi integral method and the moment of inertia method, respectively.  The  inserts in the left panels show the radial
    mode/pattern speed near the outer bar with different scales.}
\label {fig:Comp_Omega_R}
\end{figure}
\end{center}

Fig.~\ref{fig:Comp_Omega_R} shows the radial profile of the mode/pattern speed and the strength of the inner and outer bars at
selected times when (1) the two bars are aligned ($t=2.385$\,Gyr), (2) misaligned by 45 degrees ($t=2.393$\, Gyr), and
(3) perpendicular to each other ($t=2.401$\,Gyr). The different colors represent the results of the three different
methods. The magenta, blue, and black lines correspond to the Fourier method ($m=2$ mode), Jacobi integral method, and moment of
inertia method, respectively.  The mode speed of the Fourier $m=2$ mode, the pattern speed $\Omega_J$ and $\Omega_\mathrm{M}$ are
obtained using Eq.~(\ref{eq-Omega_Fourier_phase_v}), Eq.~(\ref{eq-Omega_Jacobi_LLSb}) and Eq.~(\ref{eq-Omega_MomI}), respectively.
The strength of the Fourier $m=2$ mode, $\eta_2$, and the strength $\eta_\mathrm{M}$, which is determined by the moment of inertia
method, is calculated using Eq.~(\ref{eq-Omega_Fourier_Strength}) and Eq.~(\ref{eq-lambda_MomI}).  The inset within each left panel
reveals the mode/pattern speed near the outer bar and the spiral region with an expanded scale on the $y$-axis in order to clearly
show the variation of the pattern speed in the radial direction.

It is evident from the left panels of Fig.~\ref{fig:Comp_Omega_R} that the mode/pattern speed of the two bars varies somewhat with
radius, $R$.  Specifically, the degree of variation of the mode/pattern speed of the inner bar ($R\sim 0.3-0.7$\,kpc) is greater
than that of the outer bar ($R\sim 3-5$\,kpc), especially when two bars are perpendicular to each other, as shown in
Fig.~\ref{fig:Comp_Omega_R} (e).  We note that the mode/pattern speed cannot be well determined near the galaxy center
($R<0.2$\,kpc), especially for the Fourier method (magenta line) and the moment of inertia method (black line). This is due to the
fact that these two methods require a bisymmetric component to calculate the mode/pattern speed.  Finally, none of these three
methods find similar speeds in the region between the bars, which is the expected consequence of the absence of a
  well defined steady pattern there.  For example, all three methods yield a negative pattern speed at $R=1.4 - 2$\,kpc in
Fig.~\ref{fig:Comp_Omega_R} (e), which results from the interaction of two different pattern speeds of the two bars rather than only
one single pattern.  Overall, the Jacobi integral method (blue line) reveals a smoother radial pattern speed profile than the other
two methods mainly because it probes the gravitational potential, which is relatively less reactive to the spatial density
fluctuations used in the Fourier and moment of inertia methods.

The right panels of Fig.~\ref{fig:Comp_Omega_R} compare the strength of the Fourier $m=2$ mode $\eta_2$ (magenta line) and
the strength $\eta_\mathrm{M}$ (black line) at selected times.  The strength $\eta_\mathrm{M}$ appears slightly smaller than the
strength $\eta_2$, but their variations are consistent.  The region of the inner bar and the outer bar can be easily
identified corresponding to the two peaks near $R=0.3$ and 5\,kpc, respectively. In addition, both the strength corresponding to
$\eta_2$ and $\eta_\mathrm{M}$ in the transition region lying between the inner and outer bars ($R\sim1-2$\,kpc) is small due to the
lack of a bisymmetric component.

\subsection{Time Evolution of the Mode/Pattern Speeds and Strengths}

In \S 3.1, the radial profiles of the mode/pattern speed and bar strength at selected times are displayed. In this subsection, the
temporal variation of the mode/pattern speed and the strength over the full duration of the simulation are presented.

\begin{center}
\begin{figure}[h]
  \epsscale{1.2}
\plotone{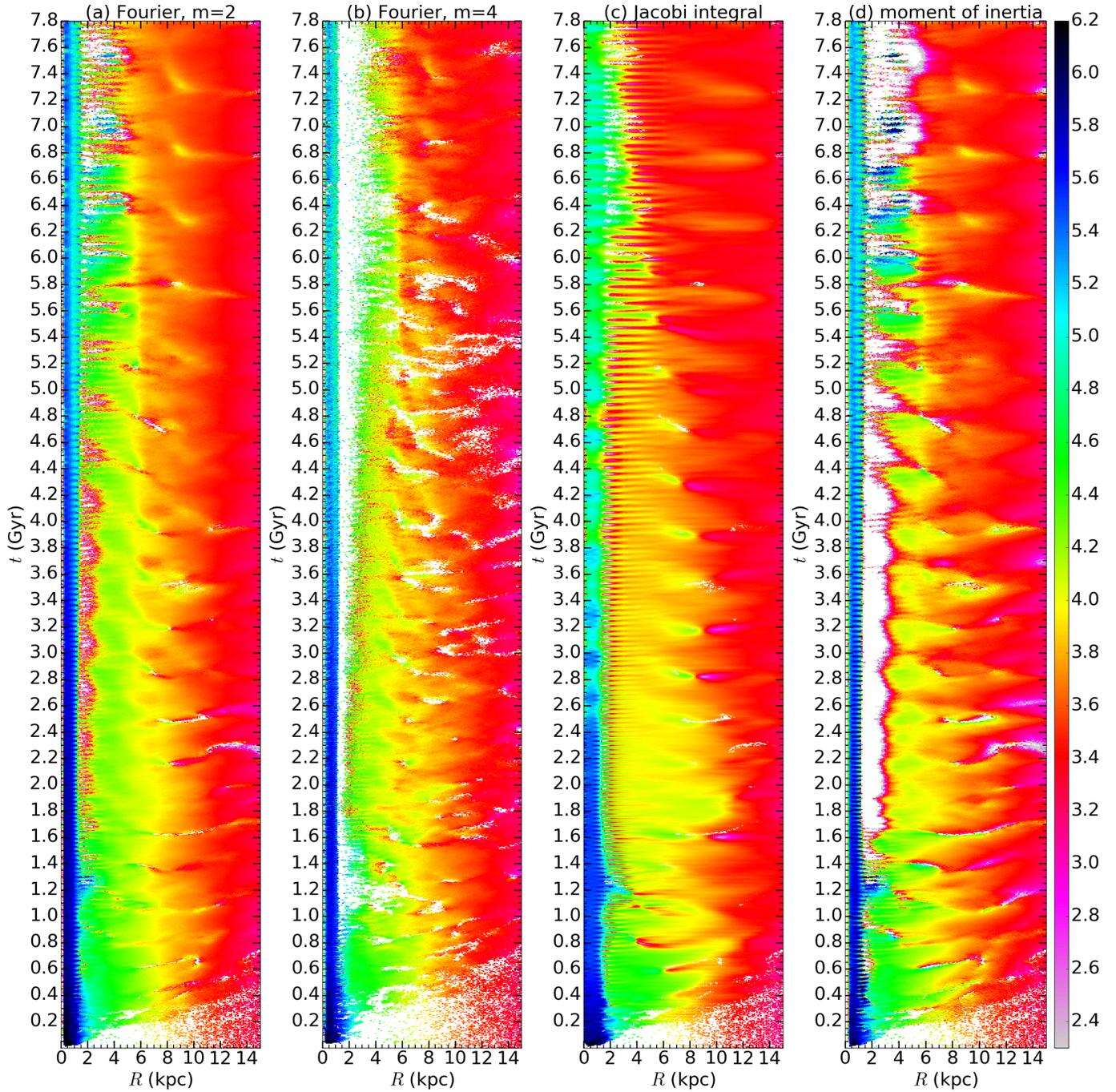}
 \caption{Time evolution of the  mode/pattern speed. The color map illustrates the  mode/pattern speed in
   asinh scale with the unit km/s/kpc. 
   The radial bin size and the time interval between two snapshots are 0.1 kpc and 1.96 Myr, respectively.
   Panels (a) and (b) present the
    mode speeds of the Fourier $m=2$ and 4 modes. Panels (c) and (d) show the pattern speeds which are obtained from the Jacobi
   integral method and the moment of inertia method, respectively.}
\label {fig:R_t_Omega_all}
\end{figure}
\end{center}

Specifically, the time evolution of the mode/pattern speed is illustrated in Fig.~\ref{fig:R_t_Omega_all}. The color map indicates
the pattern speed with the km/s/kpc unit for the asinh function argument.  The regions in white indicate the radii where the
pattern speed cannot be well determined due to a weak strength or strongly negative values.  The criteria of the strength is 0.1 for
panels (a), (c) and (d), and 0.02 for panel (b). Panels (a) and (b) illustrate the mode speed of the Fourier $m=2$ and 4 modes,
respectively. Panels (c) and (d) show the pattern speed obtained using the Jacobi integral method and the moment of inertia
method. The radial bin size is 0.1\,kpc and time interval between two snapshots is 1.96\,Myr.

As shown in Fig.~\ref{fig:R_t_Omega_all}, it is apparent that the mode or pattern speeds near the galaxy center ($R<0.2$\,kpc) and
in the region between the inner and outer bars are not well determined, as mentioned in \S 3.1.  Furthermore, the color 
  variations show that the mode/pattern speed varies not only in radius, but more importantly, in time. 

\begin{center}
\begin{figure}[h]
  \epsscale{1.2}
 \plotone{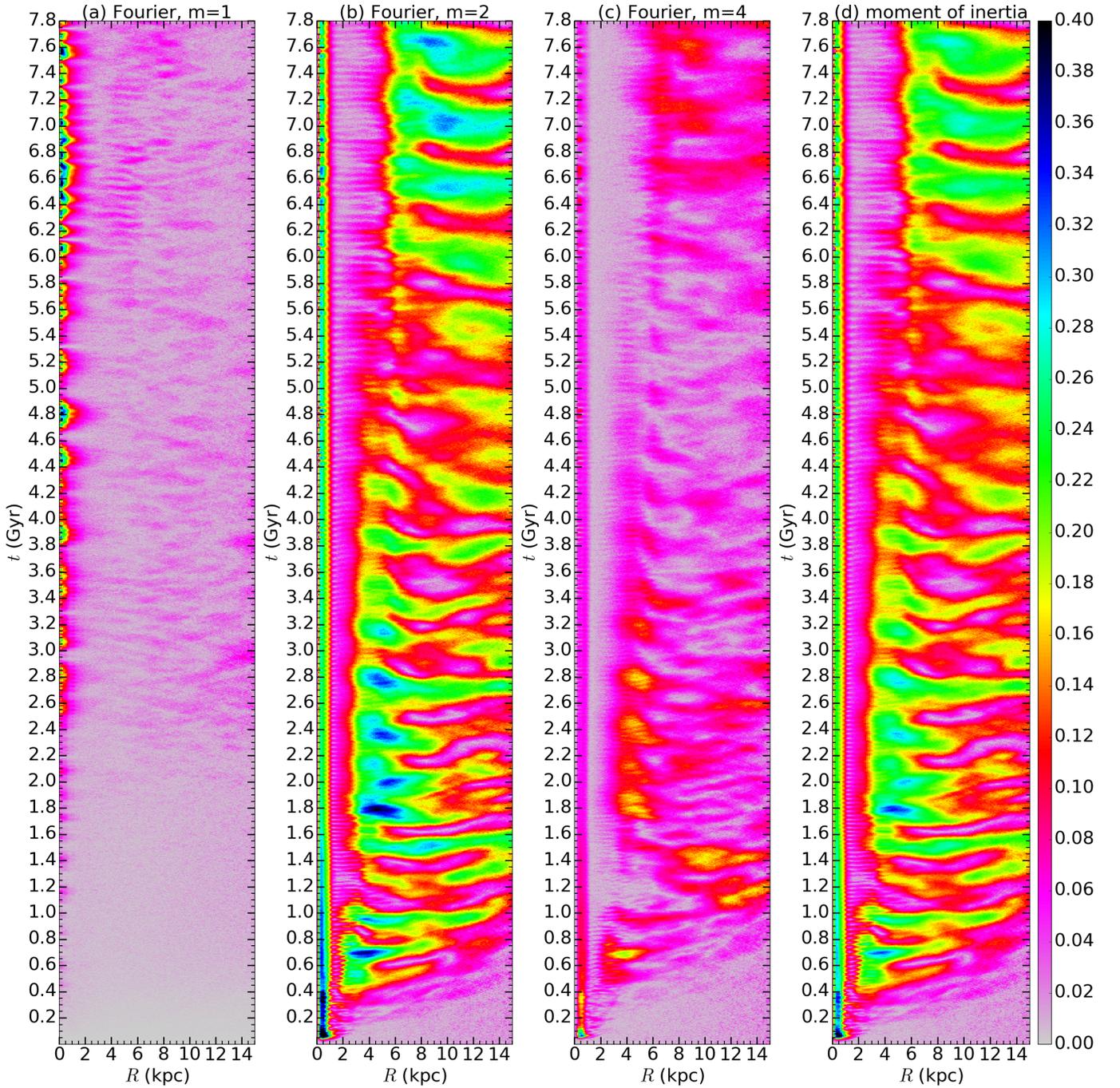}
 \caption{Time evolution of the strength. The radial bin size and time interval between two snapshots are the same in
   Fig.~\ref{fig:R_t_Omega_all}.  Panels (a), (b), and (c) show the strength of Fourier $m=1$, 2, and 4 modes. Panels (d) presents
   the strength which is determined by the moment of inertia method.}
\label {fig:R_t_Strength_all}
\end{figure}
\end{center}

The time evolution of the strength is shown in Fig.~\ref{fig:R_t_Strength_all}.  Panels (a), (b), and (c) show the strength of the
Fourier $m=1$, 2, and 4 modes. The strength of the Fourier $m=3$ mode is not shown here since it is similar to the $m=1$ mode but
weaker.  Panel (d) illustrates the strength which is determined from the moment of inertia method.  The radial bin size and time
interval between two snapshots are the same as in Fig.~\ref{fig:R_t_Omega_all}.  As can be seen from panels (b) and (d), the
strength of the Fourier $m=2$ mode is similar to the strength derived from the moment of inertia method. As for the pattern
  speed, the color variations show that the strength varies both in radius and in time. 

For example, during half of the rotation period which corresponds to the relative phase angle between the two bars changing from 0
to 90 degrees, such as from $t=2.385$ to $t=2.401$\,Gyr, the pattern speed $\Omega_J$ varies $\sim4\%$ at $R= 0.5 - 0.6\,$kpc, which
is located within the inner bar region, and the pattern speeds $\Omega_\mathrm{F}{}_2$ and $\Omega_\mathrm{M}$ change $\sim18\%$ and
$\sim37\%$, respectively.  In the same interval, the strength $\eta_2$ and $\eta_\mathrm{M}$ at $R= 0.5 - 0.6\,$kpc changes
$\sim8\%$.

Spatial variations of the pattern speed in the radial direction, for example, in the region $R= 0.3 - 0.8\,$kpc with the radial bin
size 0.1 kpc, when two bars are perpendicular to each other at $t=2.401$\,Gyr, are about $16\%$ and $12\%$ for $\Omega_J$ and
$\Omega_\mathrm{F}{}_2$ respectively. However, the variation of $\Omega_\mathrm{M}$ can be much larger and up to $75\%$.  The
strength is also seen to vary, with $\eta_2$ and $\eta_\mathrm{M}$ varying $\sim30\%$.  Hence, it is clear from
Fig.~\ref{fig:R_t_Omega_all} and Fig.~\ref{fig:R_t_Strength_all}, that both the mode/pattern speed and the strength vary
substantially with respect to the radius $R$ and time.

In comparing the different modes, the $m=2$ mode dominates the whole system for most of the time, as shown in
Fig.~\ref{fig:R_t_Strength_all} (a) -- (c). However, the strength of the Fourier $m=1$ mode increases near the galaxy center and is
comparable to the strength of the Fourier $m=2$ mode at later times such as $t=4.4$ to $t=4.5$\,Gyr.  The strength of the Fourier
$m=4$ mode also increases prior to $t=3.2$\,Gyr, but weakens thereafter.  Further details about the variation of the mode/pattern
speed and the strength which are related to the interaction between the inner bar, the outer bar and the spirals will be discussed
in \S 5.

Dynamically the strength and rotation speed variations described here are significantly strong and fast placing into
  question the often adopted assumption of individual star action conservation in galaxies (e.g.\ Sanders \& Binney 2016), at least
  in barred galaxies, and, thus,  in the Milky Way.  The overall dynamics of our simulated system is not characterized by an adiabatic slow
  evolution with low space and time frequencies, required for action conservation, but rather by a high frequency space-time
  semi-chaotic fluctuations superposed on a slow evolution. For individual stars the consequence is that they may diffuse through
  phase space with characteristic time-scales commensurable with the rotational period.  Such a behavior has already been documented
  several times in N-body simulation of barred galaxies (e.g.\ Brunetti et al.\ 2011).

\section{Time Evolution of the Two Bars}

In \S 3, we have shown that the mode/pattern speed and the strength in our model galaxy vary in radius $R$ as well as in time. In
the following, we quantitatively discuss not only the temporal variation of the pattern speeds and strengths of the inner and outer
bars, but also their sizes.

\subsection{Methods for Determining the Sizes of the Two Bars}

To determine the size of a bar, we need first to define an explicit procedure to calculate it in a reproducible way.  It was found 
that the procedures required are distinct for the inner and outer bars as their adjacent perturbing patterns are
  different. 

\subsubsection{The Inner Bar}

\begin{center}
\begin{figure}[h]
  \epsscale{1.20}
 \plotone{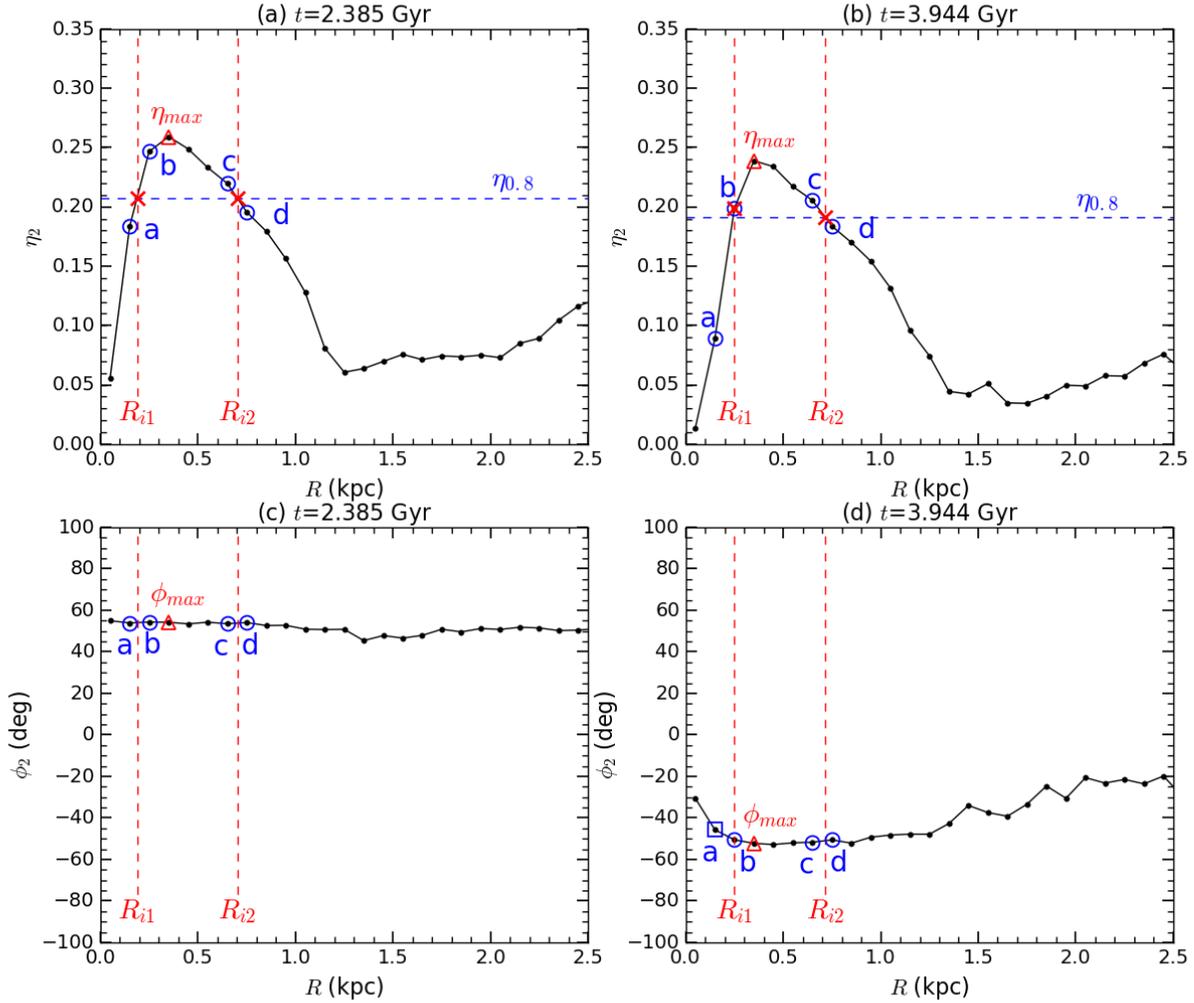}
 \caption{Two examples illustrating the procedure for determining the size of the inner bar. Panels (a) and (b) present the
   Fourier $m=2$ mode strength profile at $t= 2.385$ and 3.944\,Gyr, respectively. The red triangle marks the location of the
   maximum strength $\eta_\mathrm{max}$ at each snapshot and the blue dashed line indicates $\eta_{0.8}$, which is 80\% of
   $\eta_\mathrm{max}$.  Point 'a' -- 'd', marked by the blue circles, are used for interpolating the inner and outer radius
   $R_{i1}$ and $R_{i2}$ (marked by the red cross) as described in \S4.1.1.
   Panels (c) and (d) show
   the corresponding phase profile $\phi_2$. The red triangle ($\phi_\mathrm{max}$) indicates the phase at the radius
   corresponding to $\eta_\mathrm{max}$.  
 }
\label {fig:snap_boundary_ib}
\end{figure}
\end{center}

To determine the size of the inner bar, we consider the strength and the phase parameters from the $m=2$ Fourier mode and adopt the
following procedure.  For illustration, Fig.~\ref{fig:snap_boundary_ib} shows two examples for determining the size of the
inner bar. Panels (a) and (b) show the strength profile at $t=2.385$ and 3.944\,Gyr, respectively.  The selected particles are
located within the $|z|<$ half of the disk scale height and the radial bin size is 0.1\,kpc. Panels (c) and (d) present their
corresponding phase profile $\phi_2$.

The quantity, $\eta_{0.8}$, corresponding to a strength that is $80\%$ of its maximum value, $\eta_\mathrm{max}$, in the inner bar
region, is used to define the boundary of the bar. The $80\%$ strength allows us to exclude the regions mentioned in \S 3, where the
pattern speed is ill determined, and to cover the main portion of the bar. The blue dashed line in Fig.~\ref{fig:snap_boundary_ib}
(a) and (b) indicates $\eta_{0.8}$ at $t= 2.385$ and 3.944\,Gyr. Since the strength $\eta_2$ continuously decreases with respect to
the radial distance away from the point corresponding to $\eta_\mathrm{max}$, the radius corresponding to $\eta_{0.8}$ is derived
from the linear interpolation of the two closest points to $\eta_{0.8}$. For example, as shown in
Fig.~\ref{fig:snap_boundary_ib} (a), the point 'a' and the point 'b' are selected to interpolate for the inner radius corresponding to
$\eta_{0.8}$. The same method is used to determine the outer radius corresponding to $\eta_{0.8}$ using point 'c' and 'd'.  

A phase difference criterion is also used to determine the size of the inner bar.  The phase difference between $\phi_\mathrm{max}$,
which is the phase at the radius corresponding to $\eta_\mathrm{max}$, and the points within the inner bar is required to be less 
than $5^\circ$.  In the first example, as shown in Fig.~\ref{fig:snap_boundary_ib} (c), the phase difference between
$\phi_\mathrm{max}$ and at all points between $R=R_{a}$ and $R_{d}$ is less than $5^\circ$; hence, the boundary of the inner bar
is defined at $R_{i1}$ and $R_{i2}$.  On the other hand, in the second example, Fig.~\ref{fig:snap_boundary_ib} (d) shows that the
boundary of the inner radius $R_{i1}$ is located at point 'b' because the phase difference between $\phi_\mathrm{max}$ and point 'a'
is greater than $5^\circ$. Therefore, using the strength and the phase difference criteria together, the boundaries of the inner bar
$R_{i1}$ and $R_{i2}$, marked by the red crosses in Fig.~\ref{fig:snap_boundary_ib} (a) and (b), can be determined and the size of the inner bar is defined by the outer boundary, that is $R_{i2}$.  The radii
$R_{i1}$ and $R_{i2}$ are used for determining the strength and the mode/pattern speed of the inner bar, as described in \S
4.2.

\subsubsection{The Outer Bar}

A different method is adopted to determine the boundaries of the outer bar because the presence of the inner bar and the spiral
structure complicates the radial profiles of the strength and phase in the outer bar region.
For example, the strength of the spiral can be comparable or even stronger than the outer bar at certain times, thereby, 
making the determination of the maximum strength $\eta_\mathrm{max}$ in the outer bar region difficult.

To determine the radial boundaries of the outer bar, we adopt the following procedure.  We require that the outer bar has a
  nearly constant phase and constant pattern speed. In addition, we require that the strength within the outer bar boundaries is not
  too weak since we seek to identify a strong outer bar.  Similar to the case for the inner bar, the phase, strength, and 
  mode speed of the Fourier $m=2$ mode is considered in determining the boundaries.

To reduce the fluctuations in the phase, strength, and pattern speed, these properties are averaged using centered finite
differencing over two time-intervals of $\pm1.96$\,Myr before and after the evaluated time. We restrict the radial region to be
larger than 1\,kpc and require the simultaneous satisfaction of the  following five criteria:
\begin{enumerate}
\item The phase difference within the outer bar boundaries is less than $\pm5^\circ$;
\item The slope of the phase, that is $\frac{\phi_2(R_{i+1})- \phi_2(R_i)}{R_{i+1}-R_i}$, is less than $\pm1.5^\circ$/kpc;
\item The pattern speed difference within the outer bar boundaries is less than $\pm3$\,km/s/kpc;
\item The slope of the pattern speed, that is $\frac{\Omega_F(R_{i+1})-\Omega_F(R_i)}{R_{i+1}- R_i}$, is less than
  $\pm1$\,km/s/kpc$^2$;
\item The strength within the outer bar boundaries  exceeds the minimal strength $\eta_c=0.12$.
\end{enumerate}

With the above criteria, the boundaries of the strong outer bar, namely $R_{o1}$ and $R_{o2}$, can be determined over large time
intervals.  The outer boundary $R_{o2}$ is also used to determine the size of the outer bar .  If the above criteria are not
satisfied, we  consider the outer bar to be absent. In the following, only the strong outer bar  as defined above is discussed.
  
\subsection{The Properties of the Two Bars}

\subsubsection{The Bar Size}

\begin{center}
\begin{figure}[h]
  \epsscale{1.2}
 \plotone{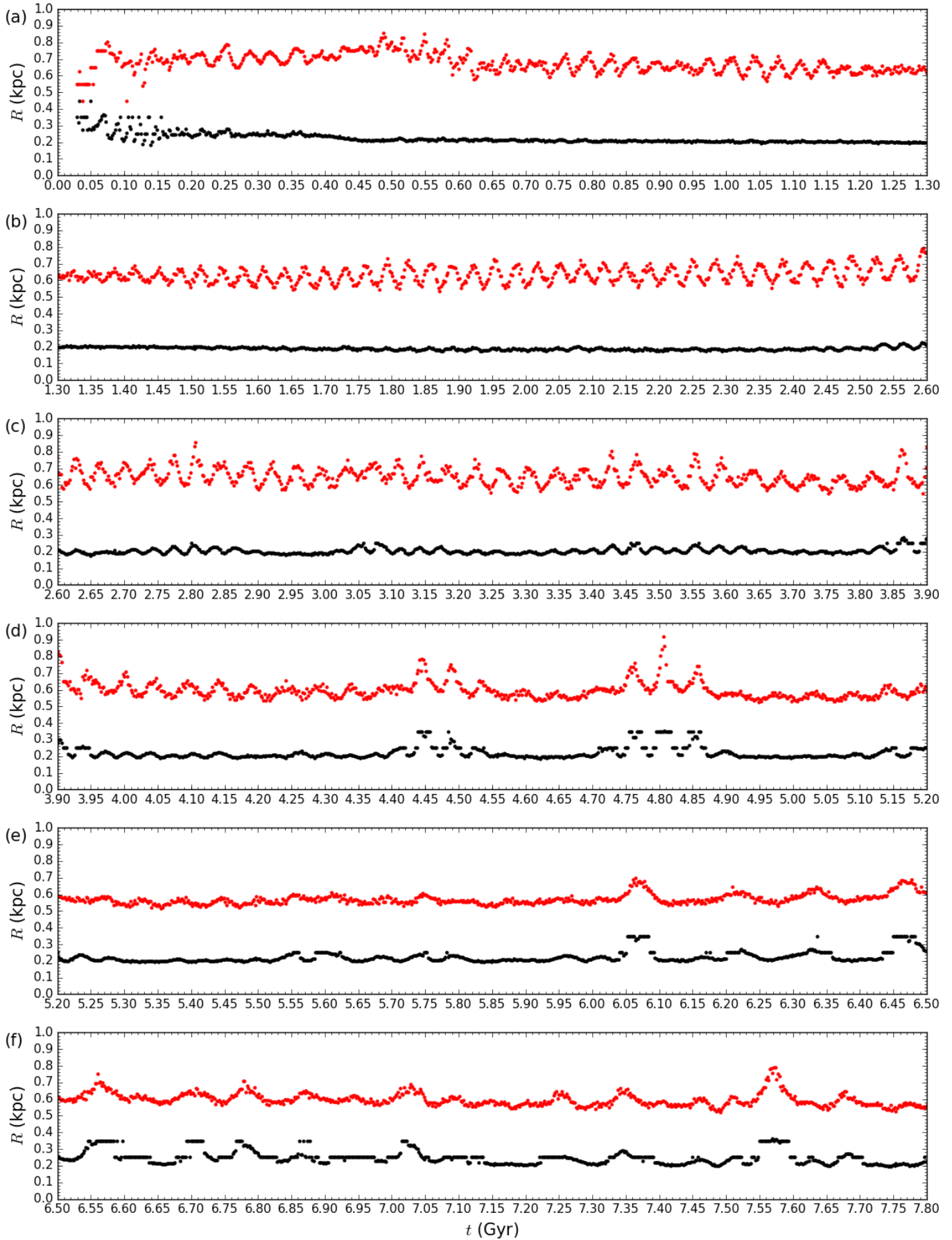}
 \caption{Time evolution of the boundaries of the inner bar. The black and red dots represent the radii $R_{i1}$ and $R_{i2}$ as
   determined in Fig.~\ref{fig:snap_boundary_ib}, respectively.}
\label {fig:t_boundary_ib}
\end{figure}
\end{center}

\begin{center}
\begin{figure}[h]
  \epsscale{1.2}
 \plotone{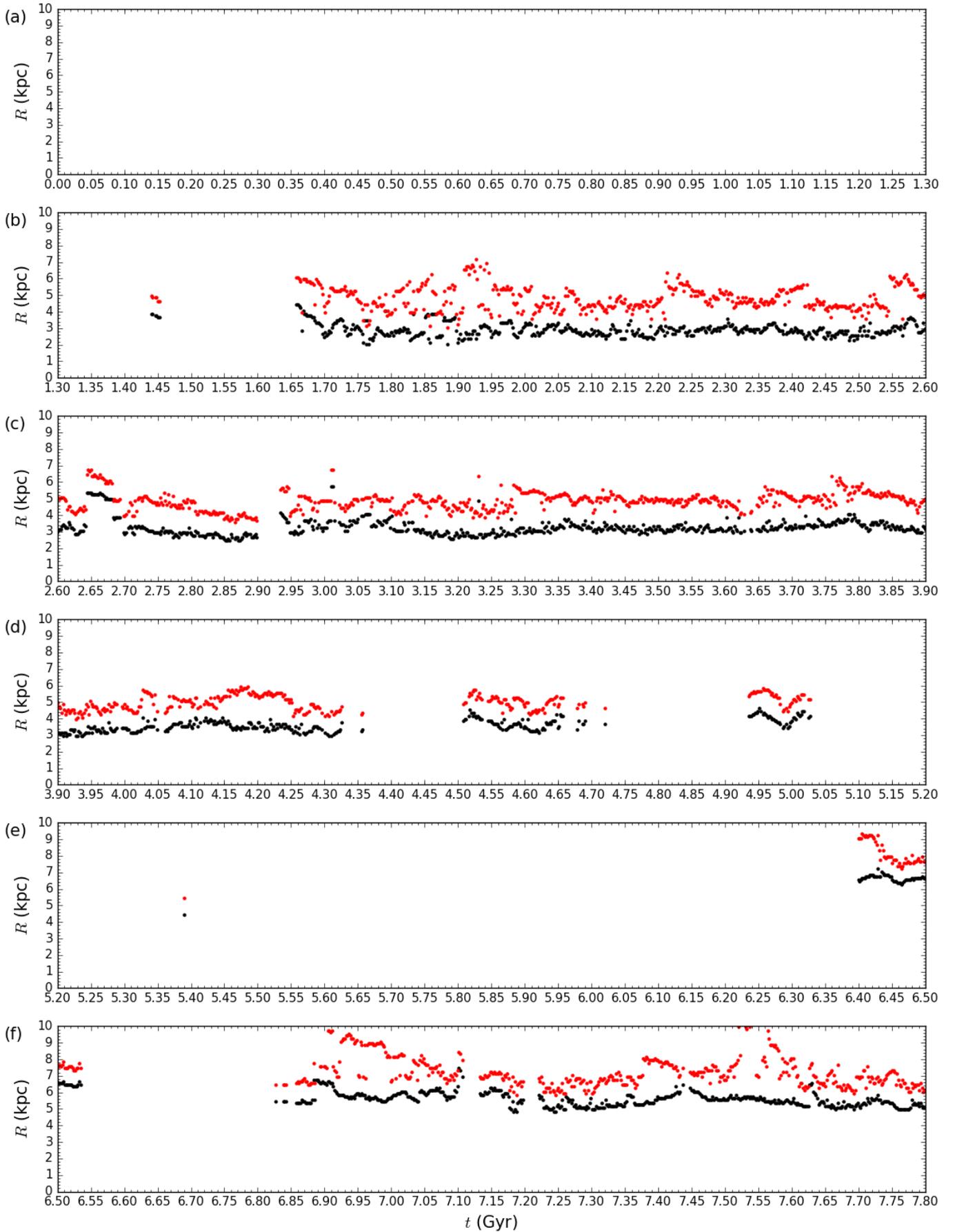}
 \caption{Time evolution of the boundaries of the outer bar. The black and the red dots represent the radius $R_{o1}$ and $R_{o2}$,
   respectively. }
\label {fig:t_boundary_ob}
\end{figure}
\end{center}

Adopting the above procedure, Fig.~\ref{fig:t_boundary_ib} and Fig.~\ref{fig:t_boundary_ob} show the time evolution of the inner
boundary (the black dots) and the outer boundary (the red dots) of the two bars once they start to develop.  As can be seen from
Fig.~\ref{fig:t_boundary_ib}, during the initial development of the inner bar ($t \sim 30$ to 70\,Myr), the size of the inner bar
increases rapidly from $\sim 0.55$ to $\sim 0.8$\,kpc.  After $t=570$\,Myr, the outer boundary $R_{i2}$ (the red dots) starts to
oscillate with the period $\sim 35$\,Myr while the inner boundary $R_{i1}$ (the black dots) does not significantly change. In
addition, $R_{i1}$ and $R_{i2}$ oscillate with greater amplitude at certain times, such as at $t\sim4.74 - 4.87$\,Gyr.

As shown in Fig.~\ref{fig:t_boundary_ob}, the outer bar starts to develop at $t\sim1.44$\,Gyr, but only lasts for $\sim10$ \,Myr
because the phase in the outer bar region changes quickly in the radial direction  which causes the phase variation to not satisfy our criterion.
At $t\sim1.66$\,Gyr, the outer bar can be defined again and lasts about 3.37\,Gyr, existing until
$t\sim5.03$\,Gyr. During this stage, the size of the outer bar is about 5\,kpc.  After $t\sim5.03$\,Gyr, the outer bar cannot be
well defined due to its weak strength for about 1.4\,Gyr.  At $t\sim6.40$\,Gyr, another outer bar starts to develop at
$R=7 - 9$\,kpc and can be well determined most of  the time after $t=6.83$ until the  simulation ends.  At some times the
  outer boundary $R_{o2}$ (red dots) suddenly increases, such as at $t=2.21$, 2.55 and 7.52\,Gyr due to the disappearance of spiral
structure (see \S 5.3).

\subsubsection{The Bar Strength}

\begin{center}
\begin{figure}[h]
  \epsscale{1.2}
 \plotone{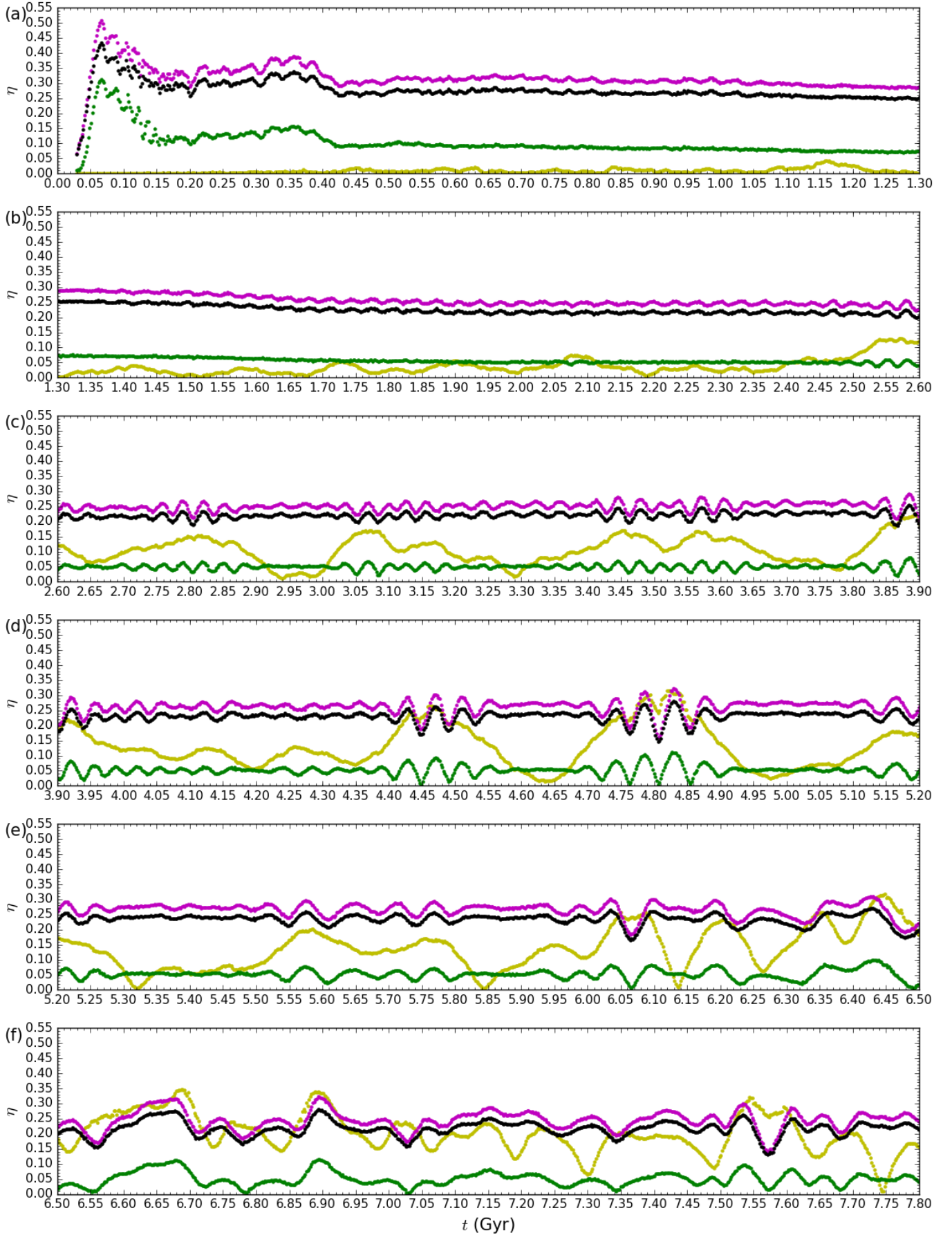}
 \caption{Time evolution of the strength of the inner bar.  The yellow, magenta and green dots illustrate the strength of the
   Fourier $m=1$, 2, and 4 mode, respectively.  The black dots shows the strength which is derived from the moment of inertia
   method.}
\label {fig:Comp_Strength_time_ib}
\end{figure}
\end{center}

\begin{center}
\begin{figure}[h]
  \epsscale{1.2}
 \plotone{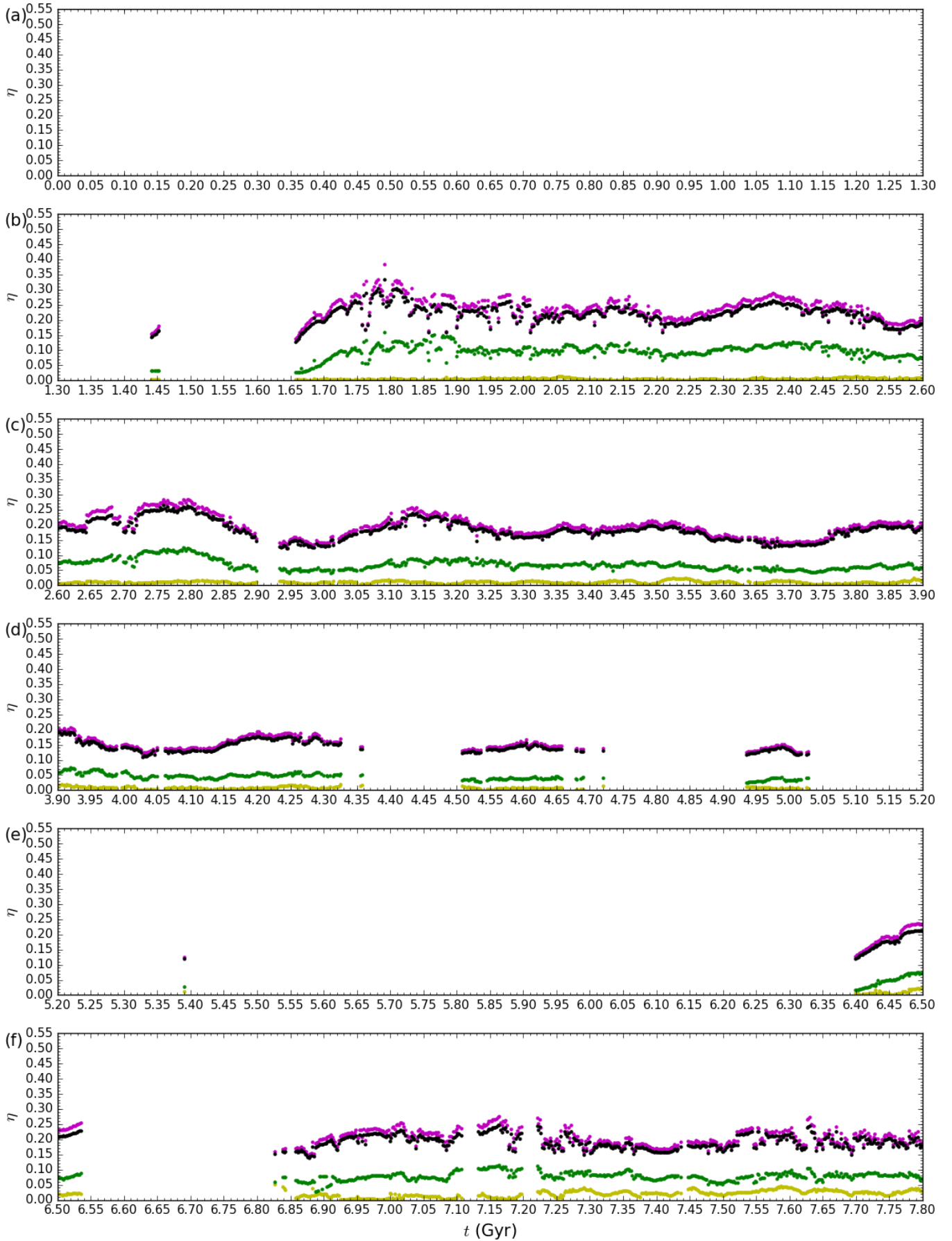}
 \caption{Time evolution of the strength of the outer bar. The dots have the same meaning as in
   Fig.~\ref{fig:Comp_Strength_time_ib}.}
\label {fig:Comp_Strength_time_ob}
\end{figure}
\end{center}

Fig.~\ref{fig:Comp_Strength_time_ib} and Fig.~\ref{fig:Comp_Strength_time_ob} show the strength of the inner and outer bar,
respectively.  The strength of the bar is calculated from the particles located between the inner and outer boundary in the
$R$-direction and $|z|<0.225$\,kpc (half of the disk scale height). The yellow, magenta and green dots illustrate the strength of
the Fourier mode $m=1$, 2, and 4. The black dots represent the strength $\eta_\mathrm{M}$.  The strength $\eta_\mathrm{M}$ is
consistent with the strength $\eta_2$ differing by only about 0.03 and 0.01 for the inner and outer bar, respectively.  The strength
$\eta_3$ is not shown here because its time evolution is similar to the strength $\eta_1$ with about half of its value.

It can be seen from Fig.~\ref{fig:Comp_Strength_time_ib} that the strength $\eta_2$ of the inner bar increases rapidly from
$\sim 0.05$ to $\sim 0.5$ during the initial development of the inner bar ($t \sim 30$ to 70\,Myr).  Subsequently, it oscillates and
decreases, reaching a relatively stable stage at $t \sim 570$\,Myr.  At $t\sim2.55$\,Gyr, the strength $\eta_1$ starts to increase
and oscillate significantly.  For example, it is about half of the strength $\eta_2$ mode at $t=2.6$\,Gyr, becoming comparable to
the strength $\eta_2$ at certain times, such as at $t\sim4.8$\,Gyr.  Compared with the strength $\eta_4$ and the strength $\eta_2$,
it can be seen that $\eta_4$ is about 5 times smaller than $\eta_2$ on average.

In the case of the outer bar, as shown in Fig.~\ref{fig:Comp_Strength_time_ob}, the $m=2$ mode dominates during the whole
simulation.  The strength $\eta_4$ is about $2.5 - 3$ times smaller than the strength $\eta_2$.  In contrast to the inner bar, the
strength $\eta_1$ is very weak and is never comparable to the strength $\eta_2$. The strength $\eta_3$ is even weaker than the
$\eta_1$ and not shown here.  Furthermore, it is evident that the strength of the outer bar oscillates in time, especially for
$\eta_2$ and $\eta_\mathrm{M}$. For example, the strength $\eta_2$ decreases at $t=2.21$, 2.55, 2.90, and 3.28\,Gyr.

\subsubsection{The  Mode/Pattern Speed}

\begin{center}
\begin{figure}[h]
  \epsscale{1.2}
 \plotone{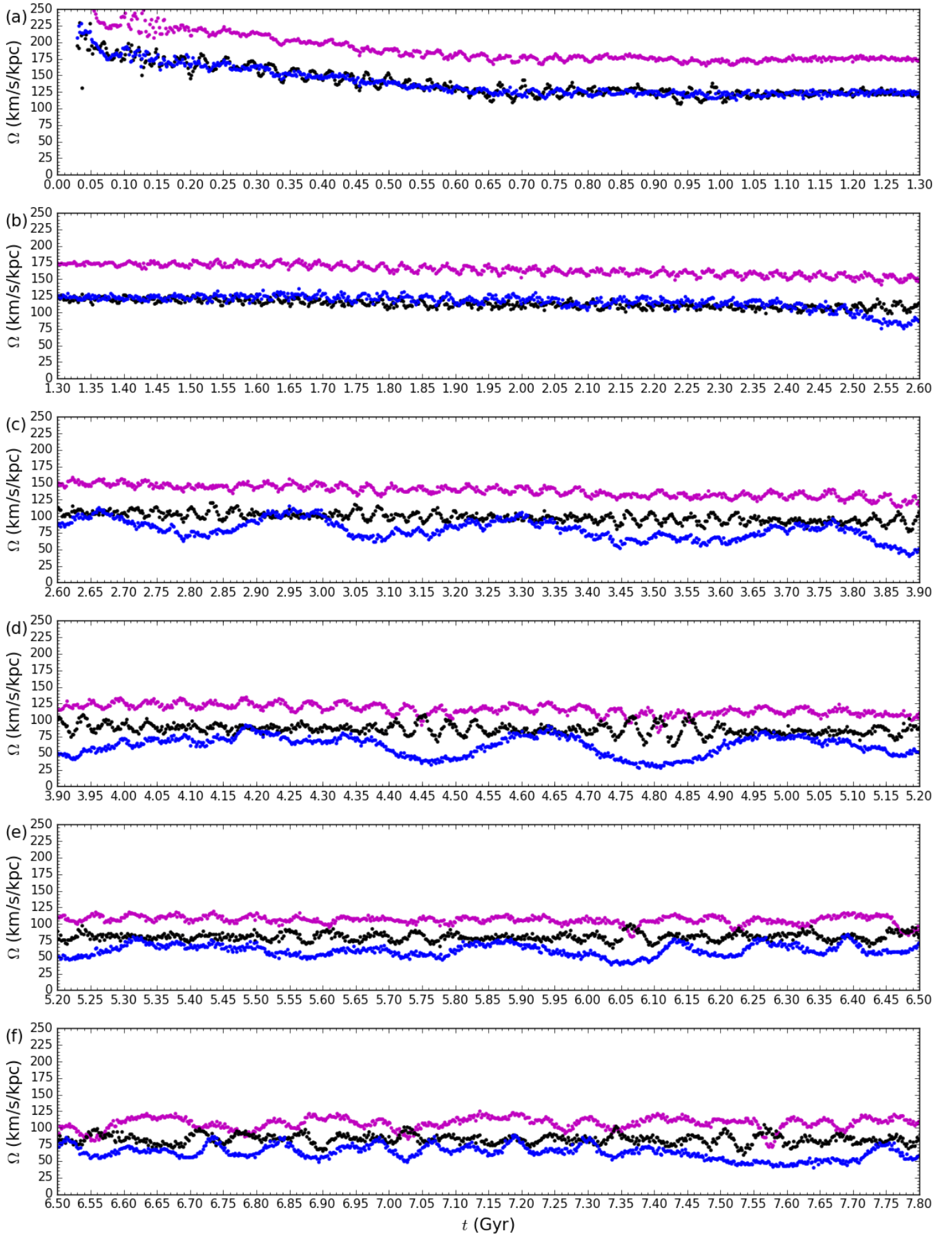}
 \caption{Time evolution of the mode/pattern speed of the inner bar.  The magenta dots represent the mode speed of the Fourier
   $m=2$ mode.  The black and blue dots illustrate the pattern speed which is determined by the Jacobi integral method and the
   moment of inertia method, respectively.}
\label {fig:Comp_Omega_time_ib}
\end{figure}
\end{center}

\begin{center}
\begin{figure}[h]
  \epsscale{1.2}
 \plotone{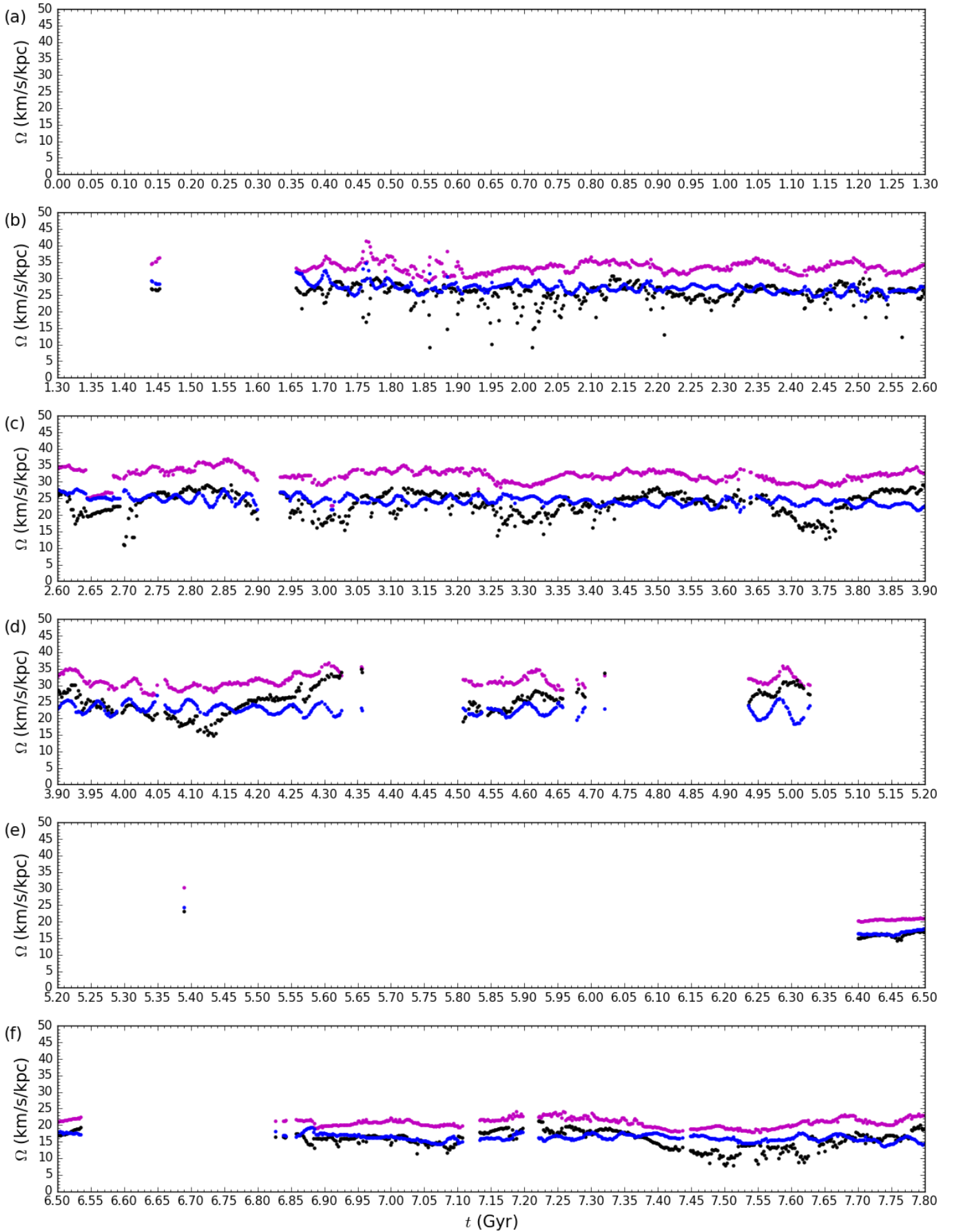}
 \caption{Time evolution of the mode/pattern speed of the outer bar. The dots have the same meaning as in
   Fig.~\ref{fig:Comp_Omega_time_ib}.}
\label {fig:Comp_Omega_time_ob}
\end{figure}
\end{center}

Fig.~\ref{fig:Comp_Omega_time_ib} and Fig.~\ref{fig:Comp_Omega_time_ob} illustrate the time evolution of the mode/pattern speed of
the inner bar and outer bar, respectively. As for calculating the strength of the bars, their mode/pattern speed is calculated only for 
the particles located between the inner and outer boundary in the $R$-direction and $|z|<$ 0.225\,kpc.  The
magenta dots represent the mode speed of the Fourier $m=2$ mode. The blue and the black dots represent the pattern speed, which is
determined by the Jacobi integral method and the moment of inertia method, respectively.

As shown in Fig.~\ref{fig:Comp_Omega_time_ib}, it is apparent that the  mode speed of the $m=2$ mode of the inner bar is
$\sim 50$\,km/s/kpc greater than the pattern speeds, which are derived from the other two methods, but has a similar evolutionary
trend.  At the beginning, the mode/pattern speed of the inner bar decreases rapidly. After the outer bar-like structure forms
($t\sim 570$\,Myr), this speed decreases gradually.

Similar to the oscillation of the boundary $R_{i2}$ as shown in Fig.~\ref{fig:t_boundary_ib}, the  mode/pattern speed also oscillates
with a higher frequency (corresponding to a time interval of $\sim 35$\,Myr).  In addition, the pattern speed $\Omega_J$ decreases
by about 50\,km/s/kpc and oscillates with a lower frequency (corresponding to a time interval of $\sim 400$ \,Myr) while the
oscillation amplitude for the pattern speed $\Omega_\mathrm{M}$ increases.

For the outer bar, as shown in Fig.~\ref{fig:Comp_Omega_time_ob}, the mode speed $\Omega_\mathrm{F}{}_2$ is about 35\,km/s/kpc and
the pattern speed $\Omega_J$ and $\Omega_\mathrm{M}$ are about 28\,km/s/kpc on average.  After $t=6.40$\,Gyr, a new longer outer bar
forms, as mentioned in \S4.2.1. Its  mode/pattern speed is about 10\,km/s/kpc lower than the shorter outer bar, which exists before
$t\sim5.03$\,Gyr.

We note that the time evolution of the mode speed $\Omega_\mathrm{F}{}_2$ has a similar oscillation trend as the pattern speed
$\Omega_\mathrm{M}$.  For example, both the  mode speed $\Omega_\mathrm{F}{}_2$ and the pattern speed $\Omega_\mathrm{M}$ decrease by
about 5\,km/s/kpc at $t=2.99$ and 3.28\,Gyr. This oscillation is consistent with the results in Fig.~\ref{fig:R_t_Omega_all} (a) and
(d). Similarly, the pattern speed $\Omega_J$ shows a high frequency ($\sim 35$\, Myr) oscillation, but with a smaller pattern speed
variation ($\sim 3$\,km/s/kpc).

\section{The Kinematic Effects in the Double-Barred Galaxy}

The description for the time evolution of the size, the strength and the mode/pattern speed of the two bars has been presented in \S
4.2.  Here, we describe the effect of the inner bar-outer bar interaction, the $m=1$ mode structure, and the outer bar-spiral
interaction on the properties of the two bars.

\subsection{The Interaction of the Two Bars}

Once the outer bar(-like) structure forms ($t=570$\,Myr), the two bars regularly interact, resulting in the oscillation in the size,
strength and mode/pattern speed of the inner bar. The period of the oscillation is related to the phase difference between the inner
and outer bars and can be calculated as follows: $\pi/(\Omega_{i}-\Omega_{o})$, where $\Omega_{i}$ and $\Omega_{o}$ are the pattern
speed of the inner and outer bar, respectively.  For example, at $t\sim2.8$\,Myr, the oscillation period is $\sim 35$\,Myr because
the difference of the pattern speed between the two bars is about 85\,km/s/kpc.

The size of the inner bar, i.e.\ the radius $R_{i2}$, is smaller when the two bars are perpendicular to each other.  This follows
from the fact that the strength of the inner bar decreases more rapidly from $\eta_{max}$ to the region between two bars
($R\sim1 - 2$\,kpc) when the two bars are perpendicular to each other, and it reaches $\eta_{0.8}$ at smaller radius, as
demonstrated in Fig.~\ref{fig:Comp_Omega_R}.  This result is in good agreement with Maciejewski \& Sparke (2000).

Similar to the oscillation of the inner bar size, as shown in Fig.~\ref{fig:Comp_Omega_time_ib}, the strength and the pattern speed
of the inner bar also oscillate corresponding to the phase angle between the inner bar and the outer bar when the $m=2$ mode
dominates the whole system.  For example, when the two bars are aligned, such as at $t\sim2.385$\,Gyr, the strength of the inner bar
 is at its local minimum and both the  mode speed $\Omega_\mathrm{F}{}_2$ and the pattern speed $\Omega_J$  are at their
  local maximum. On the other hand, when two bars are perpendicular with respect to each other, such as at $t\sim2.401$\,Gyr, the
strength of the inner bar is at its local maximum and $\Omega_\mathrm{F}{}_2$ and $\Omega_J$ are at their local
  minimum. This behavior is consistent with the results in Maciejewski \& Sparke (2000), Debattista \& Shen (2007) and Du et
al. (2015).

Due to their mutual gravitational interaction, the pattern speed of the outer bar is also affected.  As can be seen in
Fig.~\ref{fig:Comp_Omega_time_ob}, the pattern speed $\Omega_J$ of the outer bar shows a high frequency oscillation with a smaller
pattern speed variation ($\sim 3$\,km/s/kpc). When the two bars are aligned/perpendicular, $\Omega_J$ is higher/lower.  This
result also agrees with those found by Debattista \& Shen (2007) and Du et al.\ (2015).  However, in our simulation, the strength of
the outer bar is affected more by the spiral structure, which is beyond the scope of their paper, and will be mentioned in \S5.3.

\subsection{The $m=1$ Mode}

As described previously, a $m=1$ mode of the inner bar develops during the evolution.  Off-centered bars have long been
  described in observations (e.g., de Vaucouleurs \& Freeman 1970), indicating that a $m=1$ mode can occur as naturally as a $m=2$
  mode.  In our  simulation, the $m=1$ mode strength increases and oscillates significantly after $t\sim2.55$\,Gyr, as
mentioned in \S4.2.2.  At certain times, such as at $t\sim3.90$ and 4.47\,Gyr, it is even comparable to the strength of the $m=2$
mode.  When the strength $\eta_1$ becomes large, such as during $t\sim2.73 - 2.87$\,Gyr, the strength of the $m=2$ and 4 modes as
well as the size of the inner bar $R_{i2}$ start to oscillate with greater amplitude, as shown in Fig.~\ref{fig:t_boundary_ib} and
Fig.~\ref{fig:Comp_Strength_time_ib}.  This situation recurs repeatedly with the period about $300 - 400$\,Myr before
$t\sim5.85$\,Gyr.

Furthermore, the existence of the strong $m=1$ mode also affects the pattern speed of the inner bar.  $\Omega_J$ decreases and
oscillates with a lower frequency (corresponding to a time interval of $\sim 400$\,Myr), and  $\Omega_\mathrm{M}$ oscillates with
greater amplitude, as can be seen in Fig.~\ref{fig:Comp_Omega_time_ib} at $t\sim2.73 - 2.87$.

As a result, the existence of the strong $m=1$ mode not only affects the strength of other modes of the inner bar, but also the size
and the pattern speed of the inner bar. Among the three different  mode/pattern speeds, the pattern speed $\Omega_J$ is more
sensitive to the existence of the Fourier $m=1$ mode because it probes the gravitational potential, which is more reactive to the
movement of the system.

The spontaneous occurrence of odd modes here  is a reminder that it is important in N-body simulations not to impose a rigid
and fixed halo potential, as it prevents the evolution of odd modes and violates Newton's third law.

\subsection{The Spiral}

In the evolution of our galaxy model, spiral structure is found to develop in the outer parts of the galaxy.  No steady state
pattern is found to develop as the local spiral pattern speed decreases with $R$ and is actually close to the local circular
rotation frequency (Fig.~\ref{fig:Comp_Omega_R}).  As a consequence, the spiral arms are transient especially after $t\sim0.7$\,Gyr,
as mentioned in \S 2.2.  Given its location, the appearance of the spiral structure 
 and the size, strength and mode/pattern speed of the outer bar are related.  In particular, the size of the outer bar $R_{o2}$
increases upon the disappearance of the spirals, such as at $t=2.21$ and 2.55\,Gyr, or at similar phases, such as during
$t=7.52 - 7.56$\,Gyr.  Under these two circumstances, the phase near the tips of the outer bar can remain constant within
$\pm5^\circ$, the first criterion for determining the outer bar as mentioned in \S4.1.2, up to a larger radius.

\begin{center}
\begin{figure}[h]
  \epsscale{1.1}
 \plotone{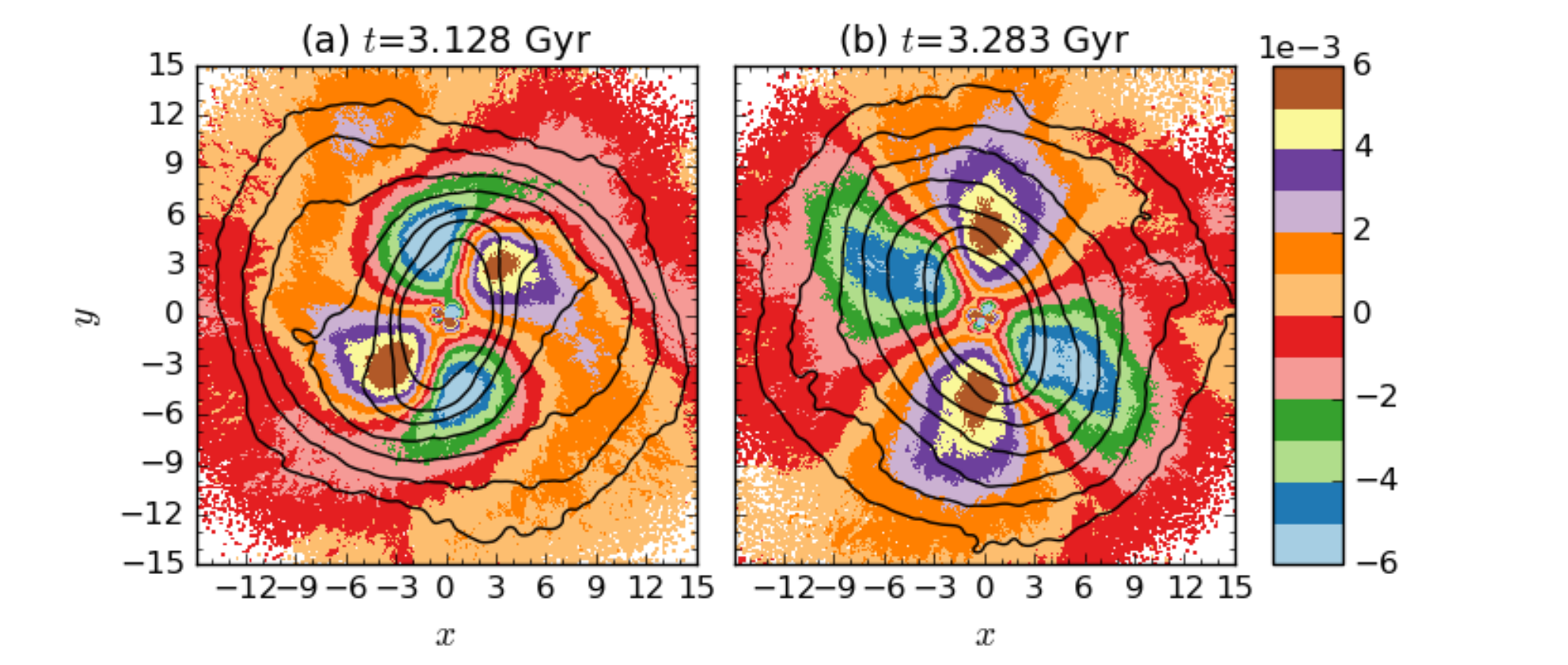} 
 \caption{ Example showing the time variation of the torque $z$-component (color) with respect to surface density contours (solid
   line).  The bin size is $(0.1\,\rm kpc)^2$.  In panel (a) the transient spiral structure
   appears, and in panel (b) disappears. The color map is the asinh of the torque scaled by a factor $10^{-3}$. The contours of the
   projected surface density are in log scale at ($-6.0$, $-5.6$, $-5.2$, $-4.8$, $-4.4$, $-4.0$, $-3.6$).}
\label {fig:Torque_asinh}
\end{figure}
\end{center}

 Fig.~\ref{fig:Torque_asinh} shows at two times the torque $z$-component when the transient spirals appear (panel (a)) and
    disappear (panel (b)). It is apparent that in the bar rotating frame, the torque changes more in the spiral region than
  in the bar region.  In other words,  the time modulation of the torque by the spirals on the bar is less noticeable than
  the torque by the bar on the spirals. Furthermore, the torque varies more in time than the morphology, as it is
    the driver of the morphology changes. However, knowing the torque is not sufficient for predicting the pattern speed
  change because a bar is not a solid body but a non-linear density wave with internal streaming.

 As mentioned in the Introduction, the cause of the time-dependence studied here could have been expected long ago from the
  work of Sellwood \& Sparke (1988), who first showed that a single bar and its adjacent spiral arms rotate in average at distinct
  speeds.  Thus, in the case that an adjacent bar or spiral structures rotate with different speeds their mutual interactions cause
  the intermediate region to be strongly time-dependent in any rotating frame. The torque modulation timescale is given by
  $\tau_\mathrm{mod} = 2\pi/\|m_{X_1}\Omega_{X_1} - m_{X_2}\Omega_{X_2}\|)$, where $m_X$ denotes the
  number of arms and $\Omega_X$ the pattern speed of structure $X$. For a typical bar--bar or a bar -- 4-arm spiral pattern, 
  $\tau_\mathrm{mod}$ is comparable (by factor $\sim 2$) to the rotational period evaluated at a radius between the two structures,
  $\tau_\mathrm{mid} = 4\pi/(\Omega_{X_1} + \Omega_{X_2})$, meaning that the modulation is dynamically fast.  
  Since the cause of time-dependence is identified for a single bar and spiral system, the description detailed here for a double bar
  system should be even more valid 
   due to the existence of additional but weaker time-dependent torque, such as the modulation of the spiral by the inner bar, and vice versa.

\section{Conclusions}

To further elaborate upon the short time scale  variations of the barred structure in a disk galaxy model reported in Wu et al.\ (2016), three
methods are used to measure the instantaneous and local mode/pattern speed rather than the time-averaged mode/pattern speed as often
done in other works. Using the Fourier mode method (which measures a mode speed), the Jacobi integral method (which measures the
potential rotation speed), and the moment of inertia method (which measures the rotation speed of the second order mass moment
tensor), we have investigated the instantaneous variations of the bar size, strength, and pattern speed of the bars.

Since the $N$-body bars and spiral arms  in our double barred galaxy model are found to be flexible at different locations, the
  different methods measure different characteristics of the structure, allowing one to quantify the time-dependence of the
structures at different locations.  The discrepancies between the methods are large when the time-dependence is strong, and small
when the pattern speed is well defined and nearly constant.  For example, we have shown that in our galaxy model, the $m=2$ Fourier
 mode speed of the inner/outer bar is larger than the pattern speed determined by the Jacobi integral method and the moment of
inertia method.  In addition, the pattern speed of the inner bar determined by the Jacobi integral method can vary by 50\,km/s/kpc
due to the appearance of the $m=1$ Fourier mode.

Although different methods yield slightly different results, all of them demonstrate that the inner and outer bars are flexible and
time-dependent in our model. As shown in Fig.~\ref{fig:Comp_Omega_R}, the pattern speed and the strength vary in radius. The
  discrepancy between the methods is especially large when the hypothesis of a single pattern at constant speed is false, for
  example between the inner and outer bar regions. The used methods allow to localize the regions with an approximate single pattern
  and quantify their time-dependence.  It is found that the regions where the patterns are well defined and constant are smaller
  than commonly assumed.

To verify the changes of the bar structure, we quantify and investigate the variations of the bar characteristics in time. The
results in \S4 confirm that both the inner and outer bars are time dependent structures. For the inner bar, the size, strength and
mode/pattern speed are affected more by the interaction with the outer bar. For example, when the two bars are aligned, the size and
 mode speed of the inner bar increases, but the strength decreases slightly. In the case of the outer bar, its characteristics are
not only affected by the inner bar but also by the adjacent spiral structure. When the spiral structure is absent, the size of the
outer bar increases, but the strength and mode/pattern speed decrease. Taken together, these results confirm that the inner and
outer bars are strongly time-dependent structures and do not have a constant mode/pattern speed. The characteristic
  time-scales of substantial pattern speed variations can be comparable to the rotational time-scale or longer, similar to the
  time-scales of substantial pattern strength variations.  These results invalidate the assumption that the actions of individual
  stars are well conserved, since the morphology changes of the potential contain space and time high-frequency semi-chaotic
  components superposed  on a slowly varying component.

In comparing the outer bar with the inner bar, the ratio of their size, strength and  mode/pattern speed are also time-dependent. The
size ratio between the outer and inner bars pulsates and increases from 7 to 10 due to the increment of the outer bar size. Because
the strength of the outer bar decreases substantially with respect to time, the strength ratio between the outer and inner bars
decreases from about 1 to 0.5 before the shorter outer bar disappears.  Finally, the  mode/pattern speed ratio between the outer and
inner bars rises from $\sim 0.23$ to 0.3 due to the decrement of mode/pattern speed of the inner bar. These results strengthen the
view that the evolution of two bars are time-dependent and do not settle into a specific resonant state.

\acknowledgments
The authors are grateful to W.~Dehnen for the gyfalcON code and D.~Yurin \& V.~Springel for the GalIC code.  The authors also thank
the support of the Theoretical Institute for Advanced Research in Astrophysics (TIARA) based in Academia Sinica Institute of
Astronomy and Astrophysics (ASIAA) and Sam Tseng for assistance on the TIARA computational facilities and resources.  The authors
would like to acknowledge the support of the Geneva Observatory and its computational facilities, as well as Yves Revaz for his
assistance.

\appendix
\section{Fourier Method}\label{sec:Fourier_Method}
\subsection{The  Mode Speed and Strength}
The Fourier method measures the  mode speed of the Fourier mode $m$ in the galactic plane.  The inverse Fourier transform of
the particle distribution can be obtained by,
\begin{equation}
  F_m = \sum_{j} m_j \exp(im\theta_j) = \sum_{j} m_j \left( \cos(m\theta_j)+i\sin(m\theta_j) \right),
  \label{eq-Omega_Fourier}
\end{equation}
where $\theta_j=\arctan(y_j, x_j)$ is the azimuthal angle of the particle $j$ given 
its cartesian coordinates $x_j$, $y_j$, and
$m_j$ is the particle mass. Therefore, the phase of any mode $m$ is
\begin{equation}
  \phi_m = \arctan(\Im(F_m) , \Re(F_m)), 
  \label{eq-Omega_Fourier_phase}
\end{equation}
where $\Re(F_m)=\sum_{j}m_j\cos(m\theta_j)$ and $\Im(F_m)=\sum_{j}m_j\sin(m\theta_j)$, and the strength of any mode $m$ is defined
as
\begin{equation}
  \eta_m = \frac{\sqrt{\Im^2(F_m)+\Re^2(F_m)}}{N}, 
  \label{eq-Omega_Fourier_Strength}
\end{equation}
where $N$ is the total number of selected particles. By differentiating the phase $\phi_m$ with respect to time, we obtain the phase
speed $\dot \phi_m$, which is $m$ times the  mode speed $\Omega_\mathrm{F}{}_m$ in ordinary space.  In compact form it reads,
\begin{equation}
  \Omega_\mathrm{F}{}_m = \frac{\dot{\phi}_m}{m} = \frac{CC_1+SS_1}{C^2+S^2}, 
  \label{eq-Omega_Fourier_phase_v}
\end{equation}
where the terms $C, S, C_1$, and $S_1$ are
\begin{eqnarray}
  C\equiv\sum_{j}m_j\cos(m\theta_j), &&  S\equiv\sum_{j}m_j\sin(m\theta_j),\\
  \label{eq-Omega_Fourier_CS_term}
  C_1\equiv\sum_{j}m_j\cos(m\theta_j)\,\dot{\theta}_j, && S_1\equiv\sum_{j}m_j\sin(m\theta_j)\,\dot{\theta}_j,
  \label{eq-Omega_Fourier_C1S1_term}
\end{eqnarray}
and $\dot{\theta}_j = (x_jv_{y,j}-y_jv_{x,j})/(x_j^2+y_j^2)$.  Thus, given the particle positions $x_j,y_j$ and 
velocities $v_{x,j},v_{y,j}$ the instantaneous  mode speed $\Omega_\mathrm{F}{}_m$ can be calculated without 
time-averaging.

\subsection{Examples}

\begin{center}
\begin{figure}[h]
  \epsscale{1.2}
 \plotone{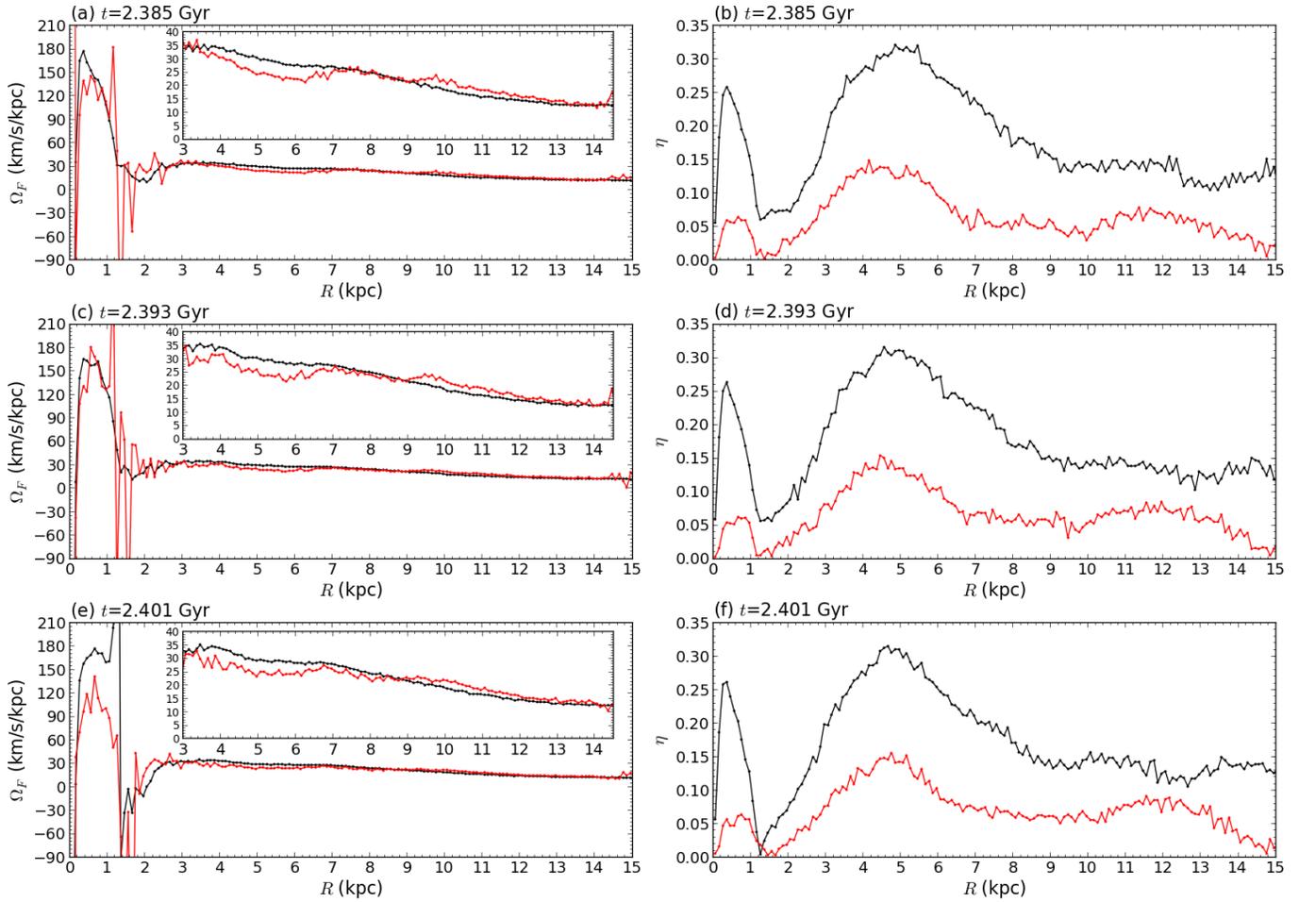}
 \caption{The radial mode speed profile (left panels) and the strength profile (right panels) which are obtained by the Fourier
   method at the selected times $t=2.385$, $2.393$, and $2.401$\,Gyr.  The inner bar and the outer bars are aligned at
   $t=2.385$\,Gyr (panels (a) and (b)), 45 degrees misalignment at $t=2.393$\,Gyr (panels (c) and (d)) and perpendicular to each
   other at $t=2.401$ \,Gyr (panels (e) and (f)).  The black line and the red line are for the $m=2$ and 4 mode, respectively. The
   subplots in the left panels show the radial mode speed near the outer bar with different scale.}
\label {fig:Omega_R_Fourier}
\end{figure}
\end{center}

In this subsection, we demonstrate the utility of the Fourier method using the selected snapshots in our double-barred system.
Fig.~\ref{fig:Omega_R_Fourier} shows examples of the radial mode speed profile (left panels) and the strength profile (right panels)
at three different selected times when the relative phase angle between the two bars is 0 ($t=2.385$\,Gyr), 45 ($t=2.393$\,Gyr) and
90 ($t=2.401$\,Gyr) degrees.  To calculate the  mode speed $\Omega_\mathrm{F}{}_m$ and the corresponding strength $\eta_m$ at
different radii, the particles are selected in an annular bin with radial and vertical sizes of 0.1\,kpc and $|z|< 0.225\,$kpc,
respectively.  The vertical bin size is half the disk scale height. The black line and the red line in
Fig.~\ref{fig:Omega_R_Fourier} denote the $m=2$ and 4 modes, which are the two dominant modes in the system at the chosen times,
respectively.  For ease of presentation in the outer region, a subplot in each left panel shows the radial  mode speed profile near
the outer bar, which has a different scale of the mode speed with respect to the inner bar.

From the left panels of Fig.~\ref{fig:Omega_R_Fourier}, it is apparent that the  mode speed of the $m=2$ and 4 modes are not
necessarily identical, and each  mode speed varies with radius within some range.  For example, as shown in panel (e), in the inner
bar region ($R\sim 0.3 - 0.7\,$kpc), the  mode speed of the $m=2$ mode is $\sim170\,$km/s/kpc, which is $\sim50\,$km/s/kpc greater
than the  mode speed of the $m=4$ mode.  Furthermore, as can be seen from the left panels, the  mode speed of the $m=2$ and $4$ mode
in the inner bar region ($R\sim 0.3 - 0.7\,$kpc) is much higher than in the outer bar region ($R\sim 3 - 5\,$kpc). For example, from
panel (a), it is seen that the  mode speed of the $m=2$ mode in the inner bar region is $\sim150\,$km/s/kpc, which is about four
times higher than in the outer bar region ($\sim 35\,$km/s/kpc).  Another significant feature shown in the left panels is that the
 mode speeds of the $m=2$ and 4 modes cannot be determined well near the galaxy center ($R<0.2\,$kpc) and also in the region between
the inner bar and the outer bar ($R\sim1 - 2\,$kpc), where the strength $\eta$ of these two modes are small and the two bars perturb
each other significantly, as shown in the right panels of Fig.~\ref{fig:Omega_R_Fourier}.

The red and the black lines in the right panels of Fig.~\ref{fig:Omega_R_Fourier} illustrate the strength of the $m=2$ and 4
modes. The two peaks of the $m=2$ strength correspond to the region of the inner bar and the outer bar, respectively. Note that the
$m=4$ mode usually appears with the $m=2$ mode. The strength of the $m=2$ mode is about 5 times larger than the $m=4$ mode for the
inner bar. On the other hand, the relative strength in the outer bar is smaller, with the strength of the $m=2$ mode about $2-3$
times larger then the $m=4$ mode. In addition, as mentioned above, at $R\sim1-2$\,kpc, which is the transition region between the
inner and outer bar, the strength of the $m=2$ mode is small, especially when two bars are perpendicular to each other, as shown in
Fig.~\ref{fig:Omega_R_Fourier} (f).

The relative angle between two bars affects the system since Fig.~\ref{fig:Omega_R_Fourier} (a), (c) and (e) show that the  mode
speed profile changes according to the phase angle between two bars, especially near the inner bar region. This might be due to the
change of the bar morphology, which causes the change of the rotation profile and the pattern speed along the major axis of the
inner bar.

In summary, the Fourier method allows a determination of the instantaneous phase, strength and  mode speed of different modes. We
find that the $m=2$ and 4 modes usually exist together with different strengths and different mode speeds for both the inner and
outer bar.  Furthermore, the determination of the strength of the $m=2$ mode can be used to determine the boundary of the bar as
described in \S 4.

\section{Jacobi Integral Method}\label{sec:Jacobi_Method}
\subsection{The Pattern Speed}

Here, we describe a method for evaluating the pattern speed of the global gravitational potential given the positions, velocities
and accelerations of a set of particles.  Under the assumptions that (1) the Hamiltonian $H=E-\bf{\Omega}_J\cdot\bf{L}$, the so
called Jacobi integral, is conserved for a test particle in a constantly rotating potential $\Phi(\mathbf{x}(t))$ (so
$\dot H = 0 = \dot E - \mathbf{\Omega}_J \dot \mathbf{L}$), and (2) the rotation axis of $\mathbf{\Omega}_J$ is along the $z$-axis.  As shown
by Pfenniger et al.\ (2018), the pattern speed $\Omega_J$ can be determined by
\begin{equation}
  \Omega_J(t)
  = \frac{\dot{E}}{\dot{L}_z}
  =\frac{\mathbf{v}(t)\cdot \mathbf{a}(t)+\dot{\Phi}(\mathbf{x}(t))}{(\mathbf{x}(t)\times\mathbf{a}(t))_z},  
\label{eq-Omega_Jacobi}
\end{equation}
where $\mathbf{x}(t)$, $\mathbf{v}(t)$, $\mathbf{a}(t)$ are the position, velocity and acceleration of a particle in the inertial frame,
respectively, and $E=\frac{1}{2}\mathbf{v}^2(t)+\Phi(\mathbf{x}(t))$, and $\dot L_z = (\mathbf{x} \times \mathbf{a})_z$ is the $z$-component of
the local torque. For our analysis, the time-derivative of the potential, $\dot{\Phi}(\mathbf{x}(t))$, is calculated by taking a finite
difference over the time-interval of $\pm 0.12$\,Myr.

In regions with a vanishing or small gravitational torque (the denominator of Eq.~(\ref{eq-Omega_Jacobi})), the pattern speed is
sensitive to noise.  A more robust approach is to solve by linear least squares the following system
\begin{equation}
  \left(\begin{array}{c} 
     \mathbf{x}_1(t)\times \mathbf{a}_1(t) \\ 
     \mathbf{x}_2(t)\times \mathbf{a}_2(t) \\ 
        \vdots \\ 
     \mathbf{x}_n(t)\times \mathbf{a}_n(t)
   \end{array}\right)_
     z {\Omega_J}(t) \approx
  \left(\begin{array}{c} 
     \mathbf{v}_1(t)\cdot \mathbf{a}_1(t) + \dot{\Phi}(\mathbf{x}_1(t)) \\ 
     \mathbf{v}_2(t)\cdot \mathbf{a}_2(t) + \dot{\Phi}(\mathbf{x}_2(t)) \\  
       \vdots \\ 
     \mathbf{v}_n(t)\cdot \mathbf{a}_n(t) + \dot{\Phi}(\mathbf{x}_n(t)) 
\end{array}\right),
\label{eq-Omega_Jacobi_LLS}
\end{equation}
simultaneously for $n$ particles belonging to a given spatial bin.  In such a system, the particles with vanishing torque do not
weight the solution, thus the result is more robust.  Indeed, the analytical least squares solution reads
\begin{equation}
 \Omega_J(t)
  = \frac{\sum_i \dot{E}_i \dot{L}_{z_i}}{\sum_i {\dot{L}_{z_i}^2}}
\label{eq-Omega_Jacobi_LLSb}
\end{equation}
which shows that the least squares solution is an arithmetic mean weighted by the torque, so small torque particles have small
weight in the mean, in contrast to solving directly Eq.~(\ref{eq-Omega_Jacobi}) for several particles and taking an average.  A
simple extension of this method described in Pfenniger et al. (2018) allows one to find the full local pattern speed \textit{vector}
by adding to the left of Eq.~(\ref{eq-Omega_Jacobi_LLS}) the $x$ and $y$ components of the torque and solving simultaneously the
least squares system for the three components of $\mathbf{\Omega}_J$.

\subsection{Examples}

The subsection below reveals the utility of the Jacobi Integral method using the same selected snapshots as in Appendix A.
Fig.~\ref{fig:Omega_xy_nosm} illustrates the pattern speed $\Omega_J$ in the $(x-y)$-plane obtained using
Eq.~(\ref{eq-Omega_Jacobi_LLSb}) for the same dimensions as in Fig.~\ref{fig:Omega_R_Fourier} (0.1\,kpc square cells in the
$(x-y)$-plane, and 0.45\,kpc height). The black curves represent the contours of the projected surface density. The upper panels
(a)--(c) and the lower panels (d)--(f) display the outer and inner bars at 15\,kpc and 2\,kpc scales, respectively.

In the upper panels, the pattern speed of the outer bar appears relatively constant ($\sim25$\,km/s/kpc) in contrast to the inner
bar, except for the regions along the semi-major and semi-minor axis of the outer bar. The pattern speed $\Omega_J$ is somewhat
noisy along the semi-major and semi-minor axis of the outer bar because the torque $\dot{L_z}$ nearly vanishes in the
swastika-shaped regions.

The lower panels of Fig.~\ref{fig:Omega_xy_nosm} illustrate the pattern speed of the inner bar. Similar to the upper panels, the
pattern speed is not well determined along the semi-major and semi-minor axis of the inner bar, where the torque is close to
zero. The morphology of the pattern speed in the $(x-y)$-plane changes according to the relative angle between the inner and the
outer bars. Specifically, the ill defined region (dark blue and grey regions) of the pattern speed is wider and circular when two
bars are perpendicular to each other, as shown in panel (f).

Fig.~\ref{fig:Omega_R_Jacobi} presents the radial distribution of the pattern speed $\Omega_J$ of the particles which are located
within the $|z|<$ half of the disk scale height, i.e.\ 0.225\,kpc, at the selected times, as in Fig.~\ref{fig:Omega_R_Fourier} and
Fig.~\ref{fig:Omega_xy_nosm}. The radial bin size is 0.1\,kpc and the color represents the number of particles in a log-scale. The
white dots represent $\Omega_J$ obtained by using Eq.~(\ref{eq-Omega_Jacobi_LLSb}) in each radial bin and the connected white line
shows the radial profile of the pattern speed $\Omega_J$.

From Fig.~\ref{fig:Omega_R_Jacobi}, it is apparent that the pattern speed $\Omega_J$ of most of the particles is about 120\,km/s/kpc
near the galaxy center ($R\sim0.5$\,kpc) and about 25\,km/s/kpc near the outer bar region ($R\sim5$\,kpc). This is consistent with
the results in Fig.~\ref{fig:Omega_xy_nosm}. We find that the pattern speed as determined from Eq.~(\ref{eq-Omega_Jacobi_LLSb}) can
be defined well near the inner bar and the outer bar region, as indicated by the white line.  However, it is difficult to define the
pattern speed close to the region between the two bars ($R \sim 1-2$\,kpc), as shown in panel (c), since the torque $\dot{L_z}$ is
close to zero and the pattern speed is not well defined in this region, as shown in Fig.~\ref{fig:Omega_xy_nosm} (f).

From the radial profile of the pattern speed (white line) in Fig.~\ref{fig:Omega_R_Jacobi}, it is evident that the pattern speed of
the inner bar ($R\sim0.3 - 0.7$\,kpc) is not constant, varying by about $\pm10$\,km/s/kpc.  However, the pattern speed of the outer
bar ($R\sim 3-5$\,kpc) is relatively constant. In summary, the Jacobi Integral method can reveal the pattern speed structure in two
dimensions, as shown in Fig.~\ref{fig:Omega_xy_nosm}, and provides another measure of the pattern speed of the inner and outer bars,
as shown in Fig.~\ref{fig:Omega_R_Jacobi}.

\section{Moment of Inertia Method}\label{sec:MoI_Method}

\subsection{The Pattern Speed and Strength}

In this subsection, we describe a method for evaluating the pattern speed of the mass weighted second moment of the particle
positions, the moment of inertia, knowing their positions and velocities. We restrict the discussion to the $x-y$ plane here, but
the method can be extended to 3D for obtaining the full pattern speed vector as described in Pfenniger et al. (2018).

The largest eigenvector of the tensor $\mathbf{I}$ of a set of particles, 
\begin{equation}
  \mathbf{I}
  \equiv 
  \left(
    \begin{array}{cc}
     I_{xx}  & I_{xy}\\
      I_{xy} &  I_{yy}
   \end{array}
  \right)
  \equiv 
  \left(
    \begin{array}{cc}
     \sum_i m_i x_i^2   & \sum_i m_i x_i y_i\\
      \sum_i m_i x_i y_i & \sum_i m_i  y_i^2
   \end{array}
  \right),
\label{eq-I_MomI}
\end{equation}
describes the principal axis of the selected particles, and allows the azimuthal angle $\phi_\mathrm{M}$ of the principal axis to be
calculated.  Therefore, if the selected particles of mass $m_i$ are located within the bar region, on the $(x-y)$-plane of Cartesian
coordinates $x_i$, $y_i$ whose center of mass is at the origin, the azimuthal angle of the semi-major axis of the bar reads,
\begin{equation}
  \phi_\mathrm{M}=
  \frac{1}{2}\mathop{\mathrm{arctan2}} \left(2I_{xy},I_{xx}-I_{yy}\right),
\end{equation}
as described in Wu et al. (2016).
The time-derivative of the angle $\phi_\mathrm{M}$ yields the angular speed of the principal axis,
\begin{equation}
  \Omega_\mathrm{M}
  \equiv
  \dot{\phi_\mathrm{M}	}(t)
  = \frac{1}{2}\frac{D_{xy}\dot{I}_{xy}-\dot{D}_{xy}I_{xy}}{D^2_{xy}+I^2_{xy}},  
\label{eq-Omega_MomI}
\end{equation}
where $D_{xy}=\frac{1}{2}\left(I_{xx}-I_{yy}\right)$ and $\dot{D}_{xy}=\sum_i m_i\left(x_iv_{xi}-y_iv_{yi}\right)$.

In addition, the strength of the bar $\eta_\mathrm{M}$ can be estimated by the eigenvalues $\lambda_\pm$ of the tensor $\mathbf{I}$.
Defining $S=\frac{1}{2}(I_{xx}+I_{yy})$, $D=\frac{1}{2}(I_{xx}-I_{yy})$ and $P=\sqrt{D^2+I_{xy}^2}$, the strength of the bar
parameter reads
\begin{equation}
  \eta_\mathrm{M}
 =1 - \sqrt{\frac{\lambda_{-}}{\lambda_{+}}},  
\label{eq-lambda_MomI}
\end{equation}
where $\lambda_{-}=S-P$ and $\lambda_{+}=S+P$. Further details can be found in Pfenniger et al. (2018).

\subsection{Examples}

The utility of the moment of inertia method will be demonstrated using three selected snapshots, the same as in Appendix A and B.
To obtain the radial profile of the pattern speed and the strength, the particles located within the $|z|<$ half of the disk scale
height are binned in $R$-direction with the bin size $\Delta R=0.1$\,kpc, the same as in the Fourier method and the Jacobi integral
method. Then, the pattern speed $\Omega_\mathrm{M}$ and the strength $\eta_\mathrm{M}$ are calculated for each annulus.

\begin{center}
\begin{figure}[h]
  \epsscale{1.2}
 \plotone{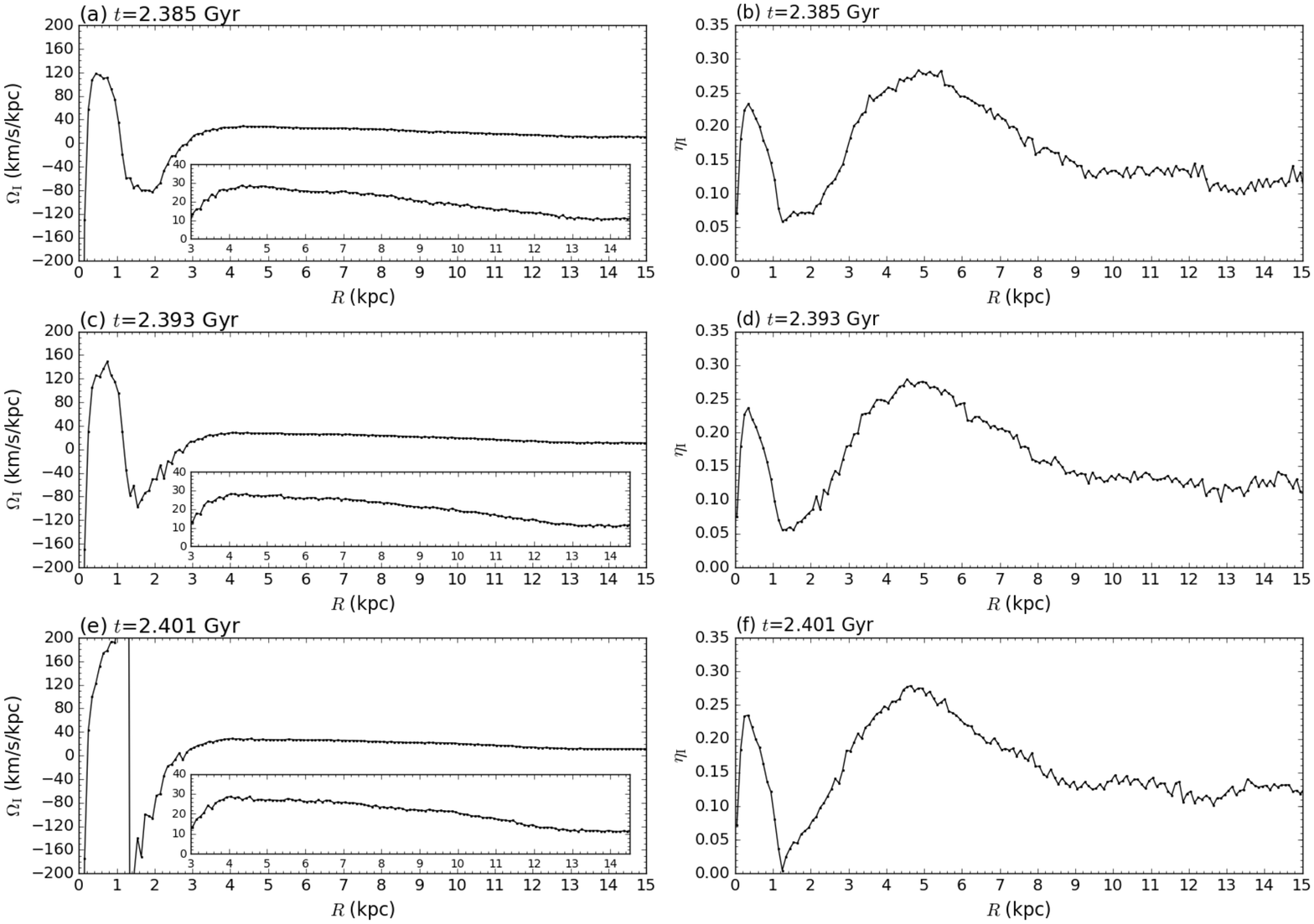}
 \caption{The radial profile of the pattern speed $\Omega_\mathrm{M}$ (left panels) and the strength $\eta_\mathrm{M}$ (right
   panels) at the selected times $t=2.385$, 2.393, and 2.401\,Gyr. The radial bin size is 0.1 kpc. The subplot in each left panel
   shows the radial pattern speed $\Omega_\mathrm{M}$ near the outer bar and spiral region with a different scale in the $y$-axis.}
\label {fig:Omega_R_MomI}
\end{figure}
\end{center}

The left panels of Fig.~\ref{fig:Omega_R_MomI} show the radial profile of the pattern speed $\Omega_\mathrm{M}$, which is obtained
using Eq.~(\ref{eq-Omega_MomI}), at selected times when the relative phase angle between two bars is 0, 45 and 90 degrees. The
subplot of each panel shows the pattern speed near the outer bar and the spiral region with a different scale in the $y$-axis to see
the changes of the pattern speed in the radial direction more clearly. From the left panels, it is evident that the pattern speed of
the two bars is not constant in radius $R$. In comparison with the pattern speed of the outer bar ($R\sim 3-5$\,kpc), the variation
of the pattern speed of the inner bar ($R\sim 0.3-0.7$\,kpc) is larger, especially when two bars are perpendicular to each other, as
shown in Fig.~\ref{fig:Omega_R_MomI} (e).  In addition, the pattern speed $\Omega_\mathrm{M}$ cannot be determined well near the
galaxy center ($R<$0.2 kpc) and in the region between two bars because the particle distribution in these regions is nearly
axisymmetric ($D_{xy} \approx 0$ and $I_{xy}\approx 0$) and the denominator of Eq.~(\ref{eq-Omega_MomI}) is about zero .

The right panels of Fig.~\ref{fig:Omega_R_MomI} illustrate the radial profile of the strength $\eta_\mathrm{M}$ at selected times.
Similar to the strength of the Fourier $m=2$ mode, two peaks around $R=0.3$ and 5\,kpc show the region of the inner bar and the
outer bar, respectively. In addition, at the transition region between the inner and outer bars ($R\sim1-2$\,kpc), the strength
$\eta_\mathrm{M}$ is small, which corresponds to the region having ill determined pattern speed.



\begin{center}
\begin{figure}[h]
  \epsscale{1.0}
 \plotone{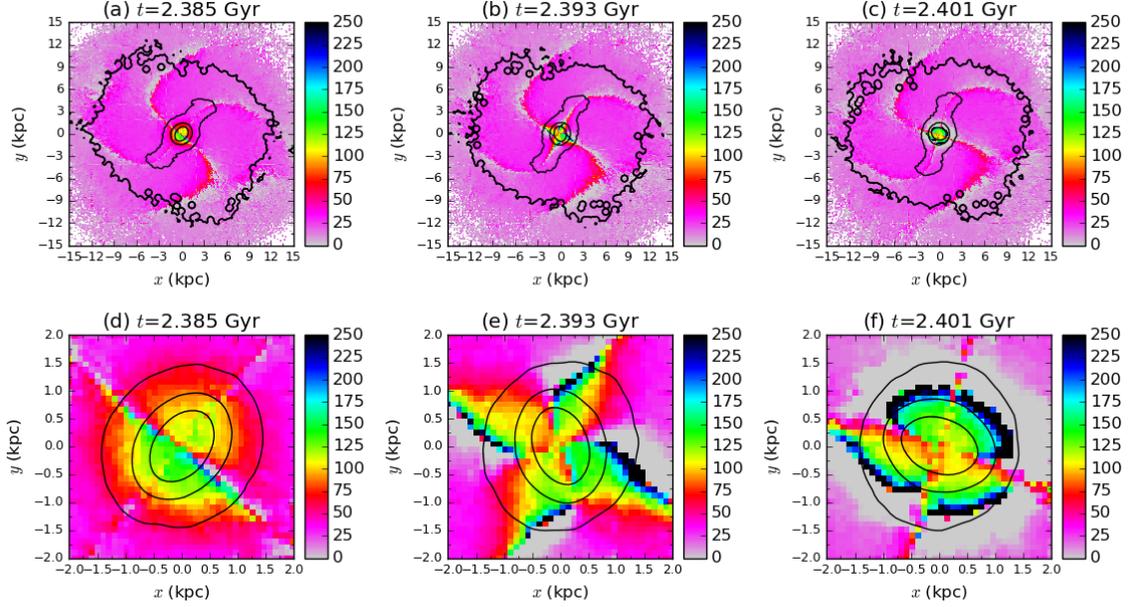}
 \caption{The pattern speed $\Omega_J$, which is determined by Eq.~(\ref{eq-Omega_Jacobi_LLSb}) in each cell, in the ($x-y$)-plane
   at selected times at 15\,kpc scale (upper panels (a)--(c)) and at 2\,kpc scale (lower panels (d)--(f)). The selected times are
   the same as in Fig.~\ref{fig:Omega_R_Fourier}.  The color map indicates the pattern speed $\Omega_J$ in unit km/s/kpc.  The black
   curves show the contours of the projected surface density in log scale, at ($-4.7$, $-3.7$, and $-2.7$) in the upper panels and
   ($-3.7$, $-2.7$, and $-1.7$) in the lower panels.  The size of each cell is 0.1\,kpc.}
\label {fig:Omega_xy_nosm}
\end{figure}
\end{center}

\begin{center}
\begin{figure}[h]
  \epsscale{0.9}
 \plotone{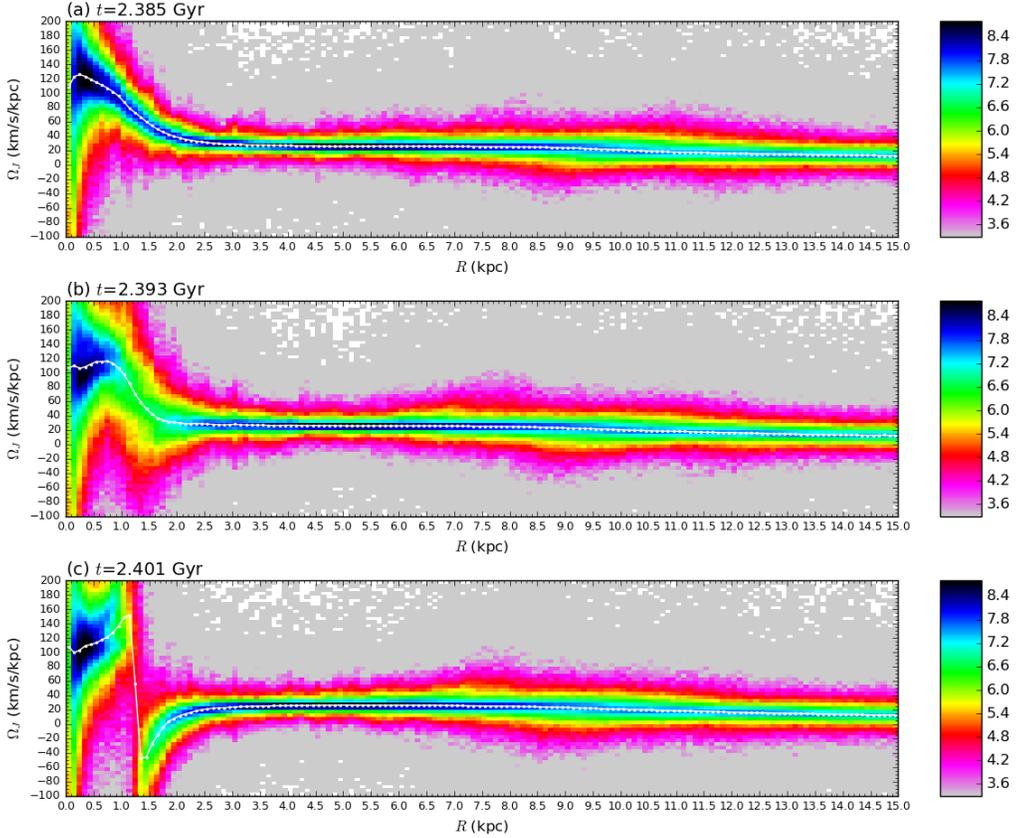}
 \caption{The radial pattern speed $\Omega_J$ profile at the selected times $t=2.385$, 2.393 and 2.401\,Gyr, the same as in
   Fig.~\ref{fig:Omega_xy_nosm} .  The color representation illustrates the number of particles in log-scale.  The radial and the
   pattern speed bin size are 0.1\,kpc and 5\,km/s/kpc.  The white dots denote the pattern speed $\Omega_J$ which is obtained by
   solving Eq.~(\ref{eq-Omega_Jacobi_LLSb}) in each radial bin, and the connected white line shows the radial pattern speed profile.
 }
\label {fig:Omega_R_Jacobi}
\end{figure}
\end{center}

\begin{thebibliography}{}
\bibitem[Aguerri(1999)]{Aguerri99} 
Aguerri, J. A. L. 1999, \aap, 351, 43

\bibitem[Athanassoula(2003)]{Atha03} 
Athanassoula, E. 2003, \mnras, 341, 1179

\bibitem[Binney \& Tremaine(2008)]{BinTre08} Binney, J. \& Tremaine,
  S. 2008, Galactic Dynamics: Second Edition, Princeton University
  Press, Princeton

\bibitem[Brunetti et al.(2011)]{BruChiPfe11}
Brunetti, M., Chiappini, C., \& Pfenniger, D. 2011, \aap, 534, A75
  
\bibitem[de Vaucouleurs at al.(1968)]{deVFre68}
  de Vaucouleurs G., de Vaucouleurs A., \& Freeman, K.~C. 1968, \mnras 139, 425

\bibitem[de Vaucouleurs \& Freeman(1970)]{deVFre70}
de Vaucouleurs, G. \& Freeman, K.~C. 1970, in The Spiral Structure of our Galaxy,
IAU Symposium 38, Becker, W. and Contopoulos, G.~I. (eds.), 356

\bibitem[Debattista \& Sellwood(2000)]{DebSel00}
 Debattista, V. P., \& Sellwood, J. A. 2000, \apj, 543, 704

\bibitem[Debattista \& Shen(2007)]{DebShe07}
 Debattista, V. P., \& Shen, J. 2007, \apj, 654, L127

\bibitem[Dehnen(2000)]{Deh00}
  Dehnen, W. 2000, \apjl, 536, L39

\bibitem[Du et al.(2015))]{DuEtal15}
  Du, M., Shen, J., \& Debattista, V. P. 2015, \apj, 804, 139

\bibitem[Du et al.(2016))]{DuEtal16}
  Du, M., Debattista, V. P., Shen, J., \& Cappellari, M. 2016, \apj, 828, 14

\bibitem[Du et al.(2017))]{DuEtal17a}
  Du, M., Shen, J., Debattista, V. P., \& de Lorenzo-C\'aceres, Adriana. 2017a, \apj, 836, 181
  
\bibitem[Du et al.(2017))]{DuEtal17b}
  Du, M., Debattista, V. P., Shen, J., Ho, Luis C., \& Erwin, Peter. 2017b, \apjl, 844, L15

\bibitem[Emsellem et al (2001)]{EmsellemEtal01}
Emsellem, E., Greusard, D., Combes, F., Friedli, D., Leon, S., P{\'e}contal, E., \& Wozniak, H., 2001, \aap, 368, 52

\bibitem[Friedli \& Martinet(1993)]{FriMar93}
Friedli, D., \& Martinet, L. 1993, \aap, 277, 27

\bibitem[Friedli \& Pfenniger (1991)]{FriPfe91}
Friedli, D., \& Pfenniger, D. 1991 
in Dynamics of Galaxies and Their Molecular Cloud Distributions, IAU Symp. 146, eds. Combes, F. and Casoli, F., 362 

\bibitem[Hasan \& Norman (1990)]{HasNor90}
Hasan, H., \& Norman, C. 1990, \apj, 361, 69

\bibitem[Hasan, Pfenniger \& Norman (1993)]{HasPfeNor93}
{Hasan}, H., \& {Pfenniger}, D., \& {Norman}, C. 1993, \apj, 409, 91

\bibitem[Hernquist \& Weinberg(1992)]{HerWei92} 
Hernquist, L., \& Weinberg, M. 1992, \apj, 400, 80

\bibitem[Maciejewski \& Sparke(2000)]{MacSpa00} 
Maciejewski, W., \& Sparke, L. S. 2000, \mnras, 313, 745

\bibitem[Marinova \& Jogee(2007)]{MarJog07} 
Marinova, I., \& Jogee, S. 2007, \apj, 659, 1176

\bibitem[Menéndez-Delmestre et al.(2007)]{MenShe07} 
Menéndez-Delmestre, K., Sheth, K., Schinnerer, E., Jarrett, T. H., \& Scoville, N. Z. 2007, \apj, 657, 790

\bibitem[Miyamoto \& Nagai(1975)]{MiyNag75}
Miyamoto, M., \& Nagai, R. 1975, \pasj, 27, 533

\bibitem[Norman, Sellwood, \& Hasan (1996)]{NorSelHas96}
Norman, C.~A., \& Sellwood, J.~A., \& Hasan, H. 1996, \apj, 462, 114

\bibitem[Pfenniger \& Norman (1990)]{PfeNor90}
Pfenniger, D., \& {Norman}, C. 1990 \apj, 363, 391

\bibitem[Pfenniger, Kanak \& Wu(2018)]{PfeSahWu18}
  Pfenniger, D., Saha, K., \& Wu, Y.-T. 2018, in preparation

\bibitem[Sanders \& Binney (2016)]{SanBin16}
Sanders, J.~L., \& Binney, J. 2016, \mnras, 457, 2107

\bibitem[Sellwood \& Athanassoula(1986)]{SelAth86}
Sellwood, J.~A. \& Athanassoula, E. 1986, \mnras, 221, 195

\bibitem[Sellwood \& Sparke(1988)]{SelSpa88}
  Sellwood, J.~A., \& Sparke, L.~S. 1988, \mnras, 231, 25P-31P
  
\bibitem[Shlosman et al.(1989)]{ShlBeg90}
Shlosman, I., Frank, J., Begelman, M.~C. 1989, \nat, 338, 455 

\bibitem[Tremaine \& Weinberg (1984)]{TreWei84}
Tremaine, S., \& Weinberg, M.~D., 1984, \apjl, 282, L5

\bibitem[Villa-Vargas et al.(2009)]{VilShl09} 
Villa-Vargas, J., Shlosman, I., \& Heller, C. 2009, \apj, 707, 218

\bibitem[Weinberg(1985)]{Wein85} 
Weinberg, M. D. 1985, \mnras, 213, 451

\bibitem[Wozniak et al.(2003)]{Wozn03}
Wozniak, H., Combes, F., Emsellem, E., \& Friedli, D.
 2003, \aap, 409, 469

\bibitem[Wu, Pfenniger \& Taam(2016)]{WuPfeTaa16}
Wu, Y.-T., Pfenniger, D. \& Taam, R. E. 2016, \apj, 830, 111

\bibitem[Yurin \& Springel(2014)]{YouSpr14}
Yurin, D., \& Springel, V. 2014, \mnras, 444, 62


\end{thebibliography}
\end{document}